\documentclass[11pt]{article}
  
\usepackage{amsmath,amssymb}

\usepackage{epsfig}  
\usepackage{graphicx}               
\usepackage{url}
\usepackage{hyperref}
\usepackage{slashed}
\usepackage[makeroom]{cancel}
\usepackage{cite}

\usepackage{pstricks}
\usepackage{color}

\usepackage{tikz}
\usetikzlibrary{positioning}
\usetikzlibrary{decorations.pathmorphing}
\usetikzlibrary{decorations.markings}
\usetikzlibrary{snakes}
 
\newcommand{\nc}{\newcommand}  

\nc{\beq}{\begin{equation}}  
\nc{\eeq}{\end{equation}}  
\nc{\beqa}{\begin{eqnarray}}  
\nc{\eeqa}{\end{eqnarray}}  
\nc{\bea}{\begin{eqnarray}}  
\nc{\eea}{\end{eqnarray}}  
\nc{\ra}{\rightarrow}  
\nc{\lsim}{\begin{array}{c}\,\sim\vspace{-21pt}\\< \end{array}}  
\nc{\gsim}{\begin{array}{c}\sim\vspace{-21pt}\\> \end{array}}  
\nc{\Tr}{{\rm Tr}}
\nc{\slsh}{\slash\hspace*{-0.22cm}}

\def\be{\begin{equation}}
\def\ee{\end{equation}}
\def\bea{\begin{eqnarray}}
\def\eea{\end{eqnarray}}
\def\bit{\begin{itemize}}
\def\eit{\end{itemize}}

\def\to{\rightarrow}

\title{  
\vspace*{-2.3cm}  
\begin{flushright}  
\normalsize{  
SLAC-PUB-15912
  }  
\end{flushright}  
\vspace{1.5cm}  
\Large  
\textbf{
Lepton Portal Dark Matter
}\vspace*{1.0cm}   
}

\author{Yang Bai\,$^{a}$ and Joshua Berger\,$^{b}$
\vspace{5mm}
\\
$^{a}$ \normalsize\emph{Department of Physics, University of Wisconsin, Madison, WI 53706, USA}  \vspace{1mm} \\
$^{b}$ \normalsize\emph{SLAC National Accelerator Laboratory, 2575 Sand Hill Road, Menlo Park, CA 94025, USA}
}

\date{}

\begin{document}  

\setcounter{page}{0}  
\maketitle  

\vspace*{1cm}  
\begin{abstract} 
We study a class of simplified dark matter models in which dark
matter couples directly with a mediator and a charged lepton. This
class of Lepton Portal dark matter models has very rich
phenomenology: it has loop generated dark matter electromagnetic
moments that generate a direct detection signal; it contributes
to indirect detection in the cosmic positron flux via dark matter
annihilation; it provides a signature of the same-flavor, 
opposite-sign dilepton plus missing transverse energy at colliders.  We
determine the current experimental constraints on the model parameter
space for Dirac fermion, Majorana fermion and complex scalar dark
matter cases of the Lepton Portal framework. We also perform a
collider study for the 14 TeV LHC reach with 100 inverse femtobarns for dark
matter parameter space. For the complex scalar dark matter case, the
LHC provides a very stringent constraint and its reach can be
interpreted as corresponding to a limit as strong as two tenths of a zeptobarn on the dark matter-nucleon scattering cross section for dark matter masses up to 500 GeV. We also demonstrate that one can improve the
current collider searches by using a Breit-Wigner like formula to fit
the dilepton MT2 tail of the dominant diboson background.
\end{abstract}  
  
\thispagestyle{empty}  
\newpage  
  
\setcounter{page}{1}

\baselineskip18pt   

\vspace{-3cm}

\section{Introduction}
\label{sec:intro}
The search for thermal relic Weakly Interacting Massive Particle
(WIMP) dark matter has a long history, particularly 
within models of weak-scale supersymmetry (SUSY)~\cite{Haber:1984rc,Jungman:1995df}.
Such models can furnish both signatures of new physics at the TeV
scale and a viable candidate for dark matter.  Collider, direct detection and
indirect detection searches for Minimal Supersymmetric Standard Model
(MSSM) dark matter particles have limited the vanilla parameter
space~\cite{CahillRowley:2012kx,Cahill-Rowley:2013dpa} and weaken the
strong tie between WIMP dark matter and the SUSY
framework.  Outside of weak-scale SUSY models, there is no specific
reason for dark matter to have mass near 100 GeV. Fortunately, the
``WIMP miracle'' provides guidance for the plausible region of dark matter mass and
interaction strength~\cite{Kolb:1990vq}.  Since even the discovery of
dark matter from multiple experimental probes is unlikely to
immediately tell us the underlying framework, in this paper we
concentrate on a class of simplified dark matter models, which serves as a
phenomenological bridge between experiments and a deep underlying
theory.  

There have been a number of recent studies of simplified dark matter models
with the emphasis on the complimentarity from different experimental
searches~\cite{Chang:2013oia,An:2013xka,Bai:2013iqa,DiFranzo:2013vra,Buchmueller:2013dya,Cheung:2013dua,Papucci:2014iwa,Simone:2014}.
Most of those studies have concentrated on dark matter interactions with
the quarks of the Standard Model (SM), which leads to a new framework
for interpretation of LHC and direct detection searches in terms of
dark matter properties. For instance, in Ref.~\cite{Bai:2013iqa} the signature of two jets plus
missing transverse energy has been studied within the context of
Quark Portal dark matter models, which is a class of simplified models
in which dark matter particles and mediators interact with a single
quark. In this paper, following our previous study in 
Ref.~\cite{Bai:2013iqa}, we concentrate on the lepton sector and study
a class of Lepton Portal dark matter models.  In these models, there
are two new particles in the dark matter sector with the
lightest one being the dark matter candidate, which must be a singlet
under electromagnetism and color.  The other particle plays the role
of mediator and connects the dark matter particle to the leptons. Obviously, to
conserve SM gauge symmetry, the mediator particle should be
charged under the electroweak symmetry.  For the dark matter
interactions to be renormalizable, the mediator must have the same
quantum numbers as the left-handed lepton weak doublet or the
right-handed charged leptons. In our study, we consider only the
latter case for simplicity.

Compared to Quark Portal dark matter models, Lepton Portal dark matter models
have totally different phenomenology at the three frontiers of the
search for WIMP dark matter.  For direct detection, unlike the Quark
Portal case, dark matter particles do not directly couple to 
target nuclei at tree level.  At one loop, the dark matter can
couple to the photon through various electromagnetic moments, which generates
the dominant interaction with the target nucleus.  The latest LUX
results from Ref.~\cite{Akerib:2013tjd} can constrain a large portion of
parameter space for Dirac fermion or complex scalar dark matter. For
indirect detection, dark matter particle annihilation can generate
electrons or positrons with a harder spectrum than the Quark Portal
case. Hence, the electron and positron flux measurement from AMS-02 in
Refs.~\cite{Aguilar:2013qda,AMS-electron-positron} becomes relevant
for the Lepton Portal models. At colliders, the Quark Portal models
have a larger signal production but also a larger QCD background. In
the Lepton Portal models, the dark matter mediator particles can be
pair produced via off-shell photons or $Z$ bosons. The
corresponding collider signature is two same-flavor charged leptons
plus missing transverse energy. Because both ATLAS and CMS
collaborations at the LHC can make very good measurements of charged
lepton momenta, the signature of dilepton plus missing transverse
energy could serve as the \emph{discovery channel} for dark matter
particles. Therefore, we pay more attention to understanding and
optimizing the key kinematic variables and work out the sensitivity at
the 14 TeV LHC.

Colliders can cover the light dark matter mass region beyond the
direct and indirect detection sensitivity. This is simply due to
different kinematics for different probes. For the three categories of
dark matter particles: Majorana fermion, Dirac fermion and complex
scalar, we have found that the 14 TeV LHC has a much better reach than
the direct detection experiments for the Majorana fermion and complex
scalar cases.  For the Majorana case, the dark matter scattering cross
section is suppressed by the dark matter velocity and predicts a very
small rate for direct detection experiments. For the complex scalar
case, the dark matter fermion partner has a large production cross
section at the LHC and a high discovery probability at the LHC.

Our paper is organized as follows. In Section \ref{sec:lepton}, we introduce
the Lepton Portal class of simplified models.  We determine the
allowed parameter space for dark matter to be a thermal relic in
Section~\ref{sec:relic-abundance}. The direct detection will be
covered in Section~\ref{sec:direct-detection}, where we perform
loop-level calculations to determine the dark matter elastic
scattering cross section. In Section~\ref{sec:indirect-detection}, we
work out constraints on model parameter space from the AMS-02 positron
and electron flux measurement. We then perform a collider study for the
sensitivity at the 14 TeV LHC with 100 fb$^{-1}$ and present summary
plots in Section~\ref{sec:collider}.  We conclude in
Section~\ref{sec:conclusion}.

\section{A Simplified Dark Matter Model: The Lepton Portal}
\label{sec:lepton}
In order for SM leptons to be a portal to the dark sector, there must
be at least two particles, one fermion and one boson, in the dark
sector.  For simplicity, we assume that there is a ${\cal Z}_2$
symmetry under which the dark sector particles are odd which
stabilizes dark matter. The lighter ${\cal Z}_2$ odd 
particle is the dark matter candidate. For the fermonic dark matter
case, we will consider both Majorana and Dirac fermions because they
have different annihilation and direct detection features.  For the bosonic dark matter case, we only consider the
complex scalar case, ignoring the real scalar case, which has
suppressed direct detection rates~\cite{Barger:2008qd}. In this paper,
we only consider the right-handed leptons as the portal particles. The
left-handed lepton case requires the dark matter partner to be a weak
doublet for renormalizable couplings and hence more degrees of
freedom. 

For fermonic (Dirac or Majorana) dark matter, $\chi$, the partner is a
scalar, $\phi$, with an electric charge $+1$. The 
renormalizable operators for the dark matter coupling to the
right-handed leptons are 
\beqa
{\cal L}_{\rm fermion} \supset \lambda_{i} \phi_{i} \overline{\chi}_L e_R^i  \,+\, {\rm h.c.} \,,
\label{eq:lag-fermion}
\eeqa
where $e^i = e, \mu, \tau$ are the charged leptons. The dark matter
mass $m_\chi$ is smaller than its partner mass $m_\phi$ such that
$\phi_{i}$ has a decay branching ratio $\mbox{Br}(\phi^i \rightarrow
\chi + \bar{e}^i)=100\%$. For a complex scalar dark matter particle, $X$,
the partner is a Dirac fermion, $\psi$, with electric charge $-1$ and the interactions
\beqa
{\cal L}_{\rm scalar} \supset \lambda_{i} X\overline{\psi^i}_L e_R^i \,+\, {\rm h.c.} \,.
\label{eq:lag-scalar}
\eeqa
Again, we have $\mbox{Br}(\psi^i \rightarrow X + e^i)=100\%$. 

To simplify our discussion, we define the Yukawa couplings in
Eqs.~(\ref{eq:lag-fermion}) and (\ref{eq:lag-scalar}) to be in the
charged-lepton mass eigenstates, so there are no new contributions to
the flavor violating processes from the dark matter sector. This
assumption can easily be arranged by implementing Minimal Flavor
Violation in the lepton sector~\cite{Batell:2013zwa}.  In the
following study, we will consider one flavor at one time.  This
assumption can easily be arranged for electron and tau coupling.  For
the muon case, it is trickier to arrange such a setup.  The results in
that case give conservative, phenomenology-based sensitivity.
Combinations of two or three flavors can be worked out based
on the results for an individual flavor. For each flavor, we have only
three parameters: the dark matter mass, its partner mass and the
coupling strength. We will work out the standard dark matter
phenomenology including thermal relic abundance, direct detection,
indirect detection and collider searches, in sequence.

\section{Relic Abundance}
\label{sec:relic-abundance}
\begin{figure}[th!]
\begin{center}
\hspace*{-0.75cm}
\begin{tabular}{c c c}
\includegraphics[width=0.34\textwidth]{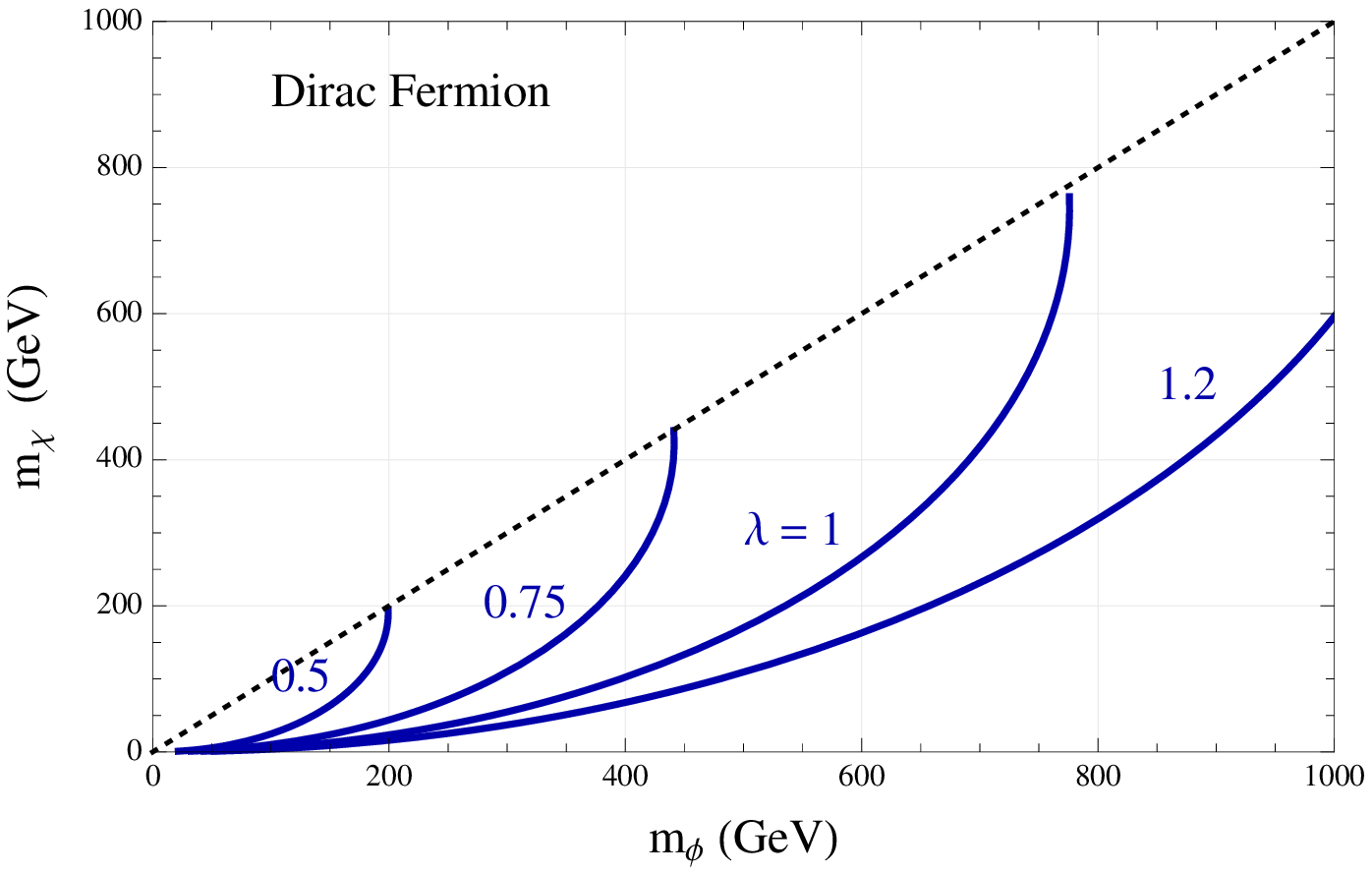} \hspace{-3mm} & 
\includegraphics[width=0.33\textwidth]{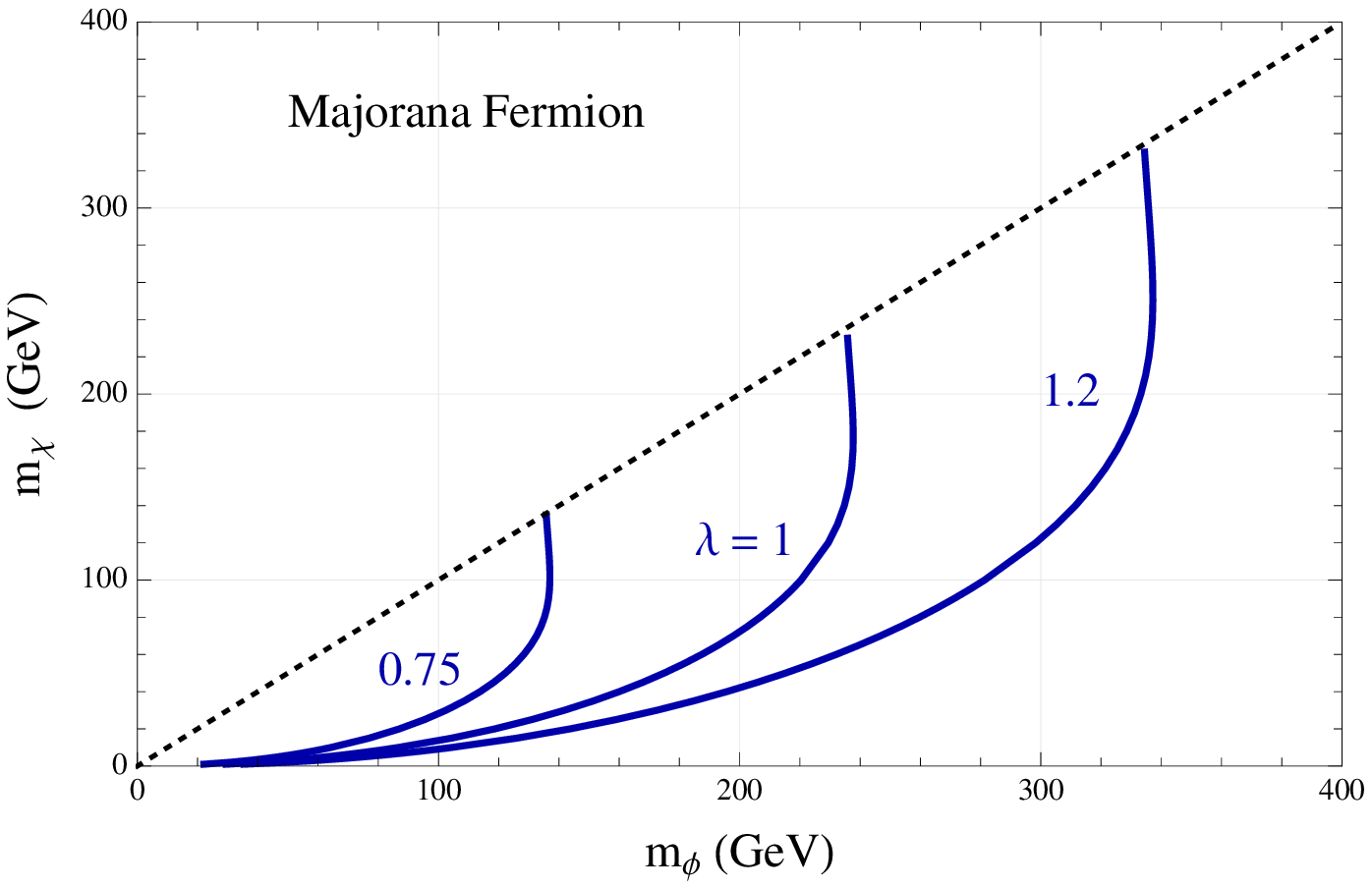} \hspace{-3mm} & 
\includegraphics[width=0.33\textwidth]{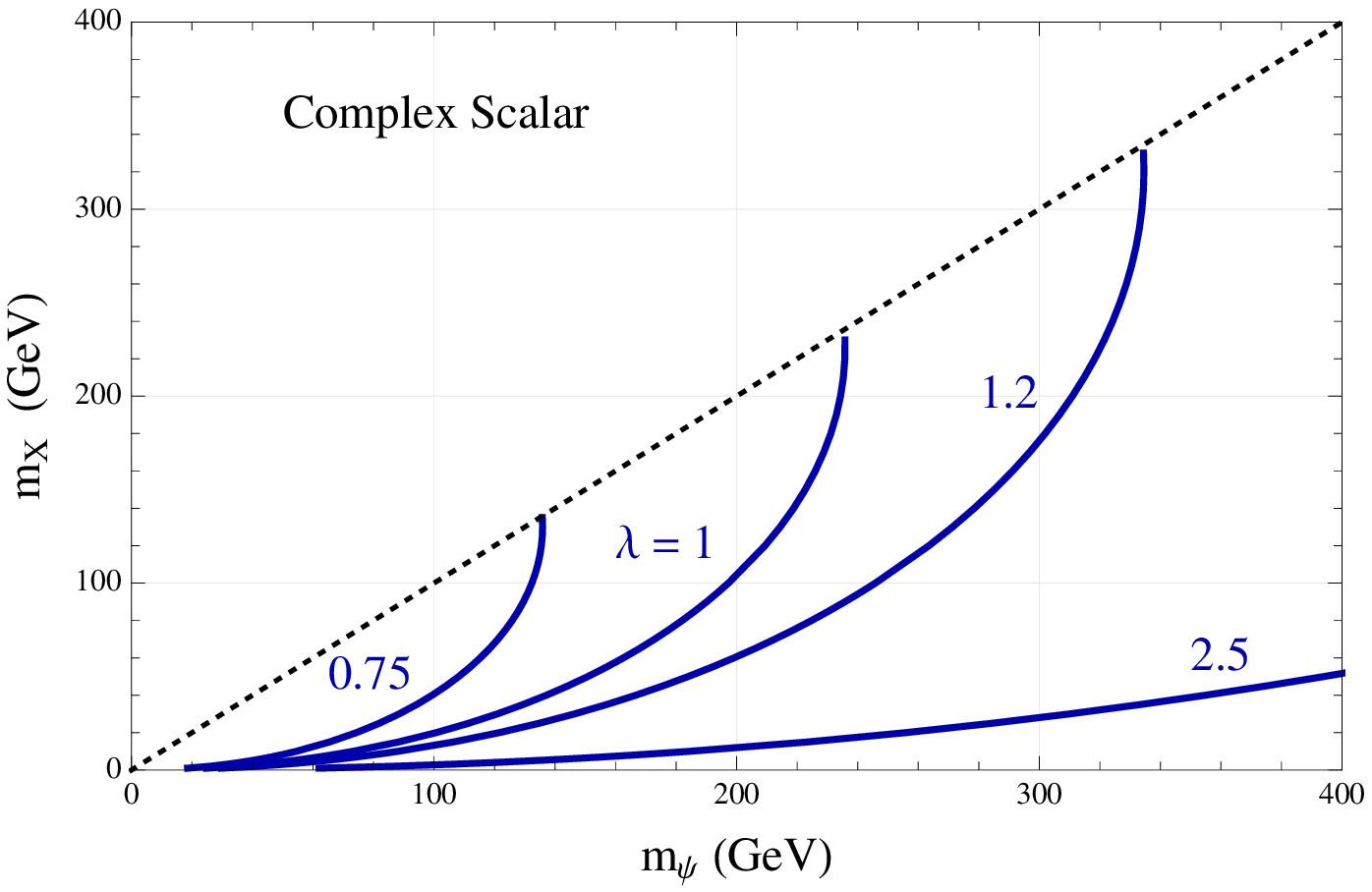} \\
{\small (a)}  & {\small (b)}  & {\small (c)}
\end{tabular}
\caption{Parameter space of a thermal dark matter for (a) Dirac fermion, (b) Majorana fermion and (c) complex scalar.}
\label{fig:relic-fermion}
\end{center}
\end{figure}

Starting with the fermonic dark matter case, the main annihilation channel is
$\overline{\chi}\chi \rightarrow \overline{e^i} e^i$ for Dirac fermion dark
matter. The dominant contribution to the annihilation cross-section is 
\beqa
\label{eq:dirac-ann}
\frac{1}{2}\,(\sigma v)^{\chi\bar\chi}_{\rm{Dirac}}  =\frac{1}{2}\left[ \frac{\lambda^4 m_\chi^2}{32\,\pi\, (m_\chi^2 + m_{\phi}^2)^2}
\,+\, v^2 \, \frac{\lambda^4\,m_\chi^2 \,( - \,5 m_\chi^4 \,-\, 18
  m_\chi^2 m_{\phi}^2 + 11 m_{\phi}^4 ) }{768\,\pi\, (m_\chi^2 +
  m_{\phi}^2)^4 } \right] \equiv s + p\,v^2  \,,
\eeqa
where $v$ is the relative velocity of two dark matter particles and is
typically $0.3\,c$ at the freeze-out temperature and $10^{-3}\,c$ at
present. We have neglected lepton masses and use $\lambda$ to
represent $\lambda_e, \lambda_\mu, \lambda_\tau$ for different
flavors. Throughout our calculation, we consider only coupling to one
flavor at a time. The factor of 1/2 in Eq.\ (\ref{eq:dirac-ann}) accounts for
the fact that Dirac dark matter is composed of both a particle and an
anti-particle. For Majorana fermion dark matter, the annihilation 
rate only contains a $p$-wave contribution at leading order in the limit of zero
lepton masses
\beqa
(\sigma v)^{\chi\chi}_{\rm{Majorana}} = v^2\,\frac{\lambda^4\,m_\chi^2\,(m_\chi^4 + m_\phi^4)}{48\pi\,(m_\chi^2 + m_\phi^2)^4} \equiv p\,v^2 \,.
\eeqa
For complex scalar dark matter, the annihilation rate of $XX^\dagger
\rightarrow \overline{e^i} e^i$ is also $p$-wave suppressed and given by
\beqa
\frac{1}{2}\,(\sigma v)^{XX^\dagger}_{\rm complex\, scalar} \,=\, \frac{1}{2}\,\left[v^2\,\frac{\lambda^4\, m_X^2}{48\,\pi\,(m_X^2 + m_\psi^2)^2} \right] \equiv p\,v^2\,.
\label{eq:complex-ann}
\eeqa
Following the same relic abundance calculation in Ref.~\cite{Bai:2013iqa}, we show the parameter space for a relic abundant dark matter for Dirac fermion, Majorana fermion and complex scalar cases in Fig.~\ref{fig:relic-fermion}. We have neglected the co-annihilation effects when the mediator and dark matter masses are degenerate (see Refs.~\cite{Ellis:1999mm, Arnowitt:2008bz} for studies on the co-annihilation region in supersymmetry models).
As one can see from Fig.~\ref{fig:relic-fermion}, the Dirac fermion
case has heavier allowed dark matter masses compared to the other two cases for a fixed value of $\lambda$.

\section{Dark Matter Direct Detection}
\label{sec:direct-detection}
\begin{figure}[th!]
\begin{center}
\hspace*{-0.75cm}
\includegraphics[width=0.6\textwidth]{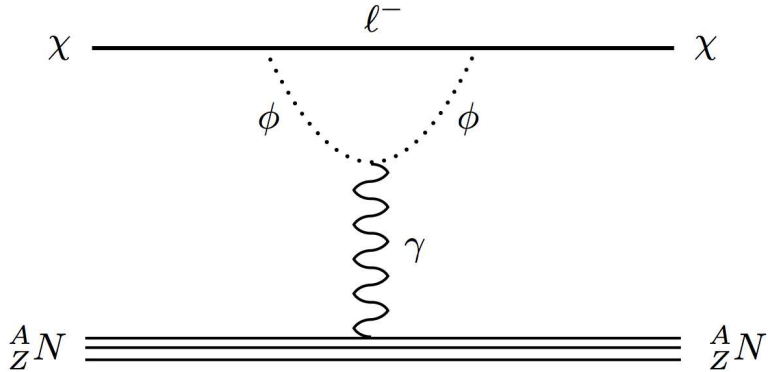}
\caption{A representative Feynman diagram for dark matter scattering off nucleus via exchanging photon at loop level. Other diagrams can have the charged lepton connect to a photon.}
\label{fig:loop-feyn}
\end{center}
\end{figure}

Since the dark matter particle only interacts with leptons at
tree-level, direct detection of dark matter in underground
experiments requires either that dark matter scatter off electrons in the
target at tree level~\cite{Kopp:2009et} or off nucleons at one-loop
level.  Because of the electron wave-function suppression, the
dominant contribution in Lepton Portal models still comes from
one-loop process with a virtual photon coupling to nucleus.  A
representative Feynman diagram is shown in Fig.~\ref{fig:loop-feyn}.

Since dark matter couples to photons at one-loop level, we will first
identify the relevant effective operators and then perform an
explicit calculation to match the coefficients of the effective
operators. To understand the physical meanings of those operators, we
will also identify the dark matter electromagnetic moments for
different operators in Appendix~\ref{app:a}.  

For the Dirac fermion case, there are two dimension-six operators
generated at one loop by which dark matter intercts with photons. They are
\beqa
{\cal O}^{\rm Dirac}_1 = \left[ \overline{\chi} \gamma^\mu ( 1- \gamma^5) \partial^\nu \chi  + \mbox{h.c.} \right] F_{\mu\nu} \,, \qquad  
{\cal O}^{\rm Dirac}_2 = \left[ i\, \overline{\chi} \gamma^\mu ( 1-
  \gamma^5) \partial^\nu \chi  + \mbox{h.c.} \right] F^{\alpha \beta}
\epsilon_{\mu\nu\alpha\beta} \,,
\label{eq:dirac-operator}
\eeqa
which yield charge-charge interactions as the leading
interactions between dark matter and nuclei~\cite{Agrawal:2011ze}.
These operators contain the charge radius, electromagnetic anapole, and
magnetic dipole moments of the Dirac dark matter.  For the Majorana
fermion case, only one chiral structure of the bi-fermion part
exists. It seems that one has two dimension-six operators at one-loop with
the forms
\beqa
{\cal O}^{\rm Majorana}_1 = \left[ - \overline{\chi}  \gamma^\mu \gamma^5\partial^\nu \chi  + \mbox{h.c.} \right] F_{\mu\nu} \,, \qquad  
{\cal O}^{\rm Majorana}_2 =  \left[ i\, \overline{\chi}  \gamma^\mu \partial^\nu \chi  + \mbox{h.c.} \right] F^{\alpha \beta} \epsilon_{\mu\nu\alpha\beta} \,.
\label{eq:majorana-operator}
\eeqa
However, one can use the Chisholm identity to prove that ${\cal
  O}^{\rm Majorana}_2=-2\, {\cal O}^{\rm Majorana}_1$ (see
Appendix~\ref{app:a} for further details)~\footnote{We thank Wai-Yee
  Keung for cross checking this point.}. Therefore, we only have a
single dimension-six operator for the Majorana fermion case. This
operator can be matched to the electromagnetic anapole 
moment of dark matter coupling to the current from the target in the
non-relativistic limit (for general discussion about anapole dark
matter see Refs.~\cite{Fitzpatrick:2010br,Ho:2012bg} and especially
Ref.~\cite{DelNobile:2014eta} for clarifying a mistake in
Ref.~\cite{Ho:2012bg}.).  

In the Lepton Portal model, the single-flavor contribution to the
effective operator in the Lagrangian is calculated and is given by
\beqa
{\cal L} \supset  c_1 {\cal O}_1 + c_2 {\cal O}_2   \,, \qquad \mbox{with} \quad
c_1 \equiv   \frac{-\lambda^2\,e}{64\pi^2 \, m_\phi^2}   \left[ \frac{1}{2} + \frac{2}{3} \ln\left( \frac{m^2_{e^i}}{m^2_\phi} \right)\right] \,, \quad c_2 \equiv   \frac{-\lambda^2\,e}{64\pi^2 \, m_\phi^2} \frac{1}{4} 
 \,,
\eeqa
for both Dirac and Majorana cases~\footnote{We have checked our
  formulas against Ref.~\cite{Agrawal:2011ze} and agree with their
  calculation.}. For muon and tau cases, we use the masses
for $m_{e^i}$. For the electron case, for which the lepton mass is below the
exchange momentum of the scattering process, one should replace
$m_{e^i}$ by the exchange momentum $|\vec{q}|$ with $\vec{q}^2 =2
\mu^2_{\chi T} v^2 (1-\cos{\theta}) = {\cal O}(10-100)$~MeV depending
on the dark matter mass. Here, $\mu_{\chi T}$ is the reduced mass of
the dark matter-nucleus system; $\theta$ is the scattering angle in
the center-of-mass frame; $v$ is the dark matter velocity in the lab
frame.  

For the Dirac dark matter case, neglecting the form factors on the
dark matter side, we still have two different moments for photon
coupling to the nucleus in the target. Since the charge and magnetic
dipole moment parts have different form factors, we keep track of
those parts in our calculation. For the spin-independent coupling to
the charge of the nucleus, the differential scattering cross
section in the recoil energy, $E_R=|\vec{q}|^2/2m_T$, at the leading
order in $v^2$ is
\beqa
\frac{d\sigma^E_T}{d E_R} =\left[ c_1^2\,e^2 \,Z^2\, \frac{m_T}{2\pi v^2}  
+ c_2^2\,e^2\,Z^2\, \left( \frac{4m_\chi^2}{\pi E_R} - \frac{2 m_\chi^2 m_T}{\pi \mu^2_{\chi T}v^2} \right) \right] F_E^2(q^2) \,,
\label{eq:dirac-charge}
\eeqa
 where $Z$ is the charge of the target nucleus and $F_{E}(q^2)$ is the
 electric form factor of the target nucleus~\footnote{Our result is
   different from Ref.~\cite{Agrawal:2011ze}. We don't have a term
   proportional $c_1 c_2$ because the dark matter (a point-like
   particle) charge and magnetic-dipole moment parts should be summed
   together in the matrix element calculation and their $c_1c_2$ terms
   cancel each other.}. For the coupling to the nuclear
 magnetic dipole moment, we obtain a differential cross section
 \beqa
 \frac{d\sigma^M_T}{d E_R} = c_2^2 \,e^2\,\frac{4}{\pi} \frac{m_\chi^2}{m_T v^2}\, \frac{m_T^2 \lambda_T^2}{m_N^2 \lambda^2_N} \frac{J_T+1}{3 J_T} F_M^2(q^2) \,.
 \label{eq:dirac-dipole}
 \eeqa
Here, $\lambda_N = e/2 m_N$ is the nuclear magneton; $m_N$ is the
nucleon mass; $\lambda_T$ is the target nucleus magnetic moment; $J_T$
is the spin of the target nucleus; $F_M(q^2)$ is the form factor of
the nucleus magnetic dipole moment. For the Xenon element, the two
most abundant and stable isotopes have
$\lambda_T/\lambda_N(^{129}_{54}\mbox{Xe})=-0.778$ with
$J_T(^{129}_{54}\mbox{Xe})=1/2$ and an abundance of 26.40\% and
$\lambda_T/\lambda_N(^{131}_{54}\mbox{Xe})=+0.692$ with
$J_T(^{131}_{54}\mbox{Xe})=3/2$ and an abundance of
21.23\%~\cite{Raghavan:1989zz} (see Ref.~\cite{Banks:2010eh} for a
collection of nuclear magnetic moments for more elements in direct
detection experiments). Comparing
Eqs.~(\ref{eq:dirac-charge})(\ref{eq:dirac-dipole}), one can see that
the magnetic moment part is sub-leading compared to the charge part
for a light dark matter because of the $m_\chi/m_T$ and $|c_2/c_1|\sim
1/40(1/20)$ suppression factors for muon(tau). Keeping the leading and
first term in Eq.~(\ref{eq:dirac-charge}), we have the same $v^2$
dependence as the spin-independent scattering.  We obtain
approximate results for dark matter-nucleus and dark matter-nucleon
scattering cross sections as
\beqa
\sigma_{\chi T} =c_1^2\,e^2 \,Z^2\, \frac{\mu_{\chi T}^2}{\pi }  \,, \qquad 
\sigma_{\chi N} =c_1^2\,e^2 \,Z^2\, \frac{\mu_{\chi N}^2}{A^2\,\pi }  \,,
\eeqa
where $\mu_{\chi N}$ is the reduced mass of the dark matter-nucleon
system. Using the LUX result~\cite{Akerib:2013tjd}, we show the
constrains on the model parameter space in
Fig.~\ref{fig:constraint-dirac} using $Z=54$ and $A=129$. 

For the Majorana fermion case, the dimension-six operators in
Eq.~(\ref{eq:majorana-operator}) couple to the charge and the magnetic
dipole moment of the nucleus. The differential cross section in $E_R$
is suppressed by an additional power of $v^2$ compared to the Dirac
fermion case and leads a weak direct detection signals. The form for
the charge part is
\beqa
\frac{d\sigma^E_T}{d E_R} = (c_1-2 c_2)^2\,e^2 \,Z^2\, \frac{m_T}{4\pi}  \left( 2 - \frac{m_T\,E_R}{\mu_{\chi T}^2 v^2} \right)  F_E^2(q^2) \,.
\label{eq:majorana-charge}
\eeqa
The dipole moment part has
\beqa
 \frac{d\sigma^M_T}{d E_R} =(c_1-2c_2)^2\,e^2 \,\frac{1}{2\pi} \frac{E_R}{v^2}\, \frac{m_T^2 \lambda_T^2}{m_N^2 \lambda^2_N} \frac{J_T+1}{3 J_T} F_M^2(q^2) \,,
 \label{eq:majorana-dipole}
 \eeqa
which agrees with the results in Ref.~\cite{DelNobile:2014eta} and
disagrees with Ref.~\cite{Ho:2012bg}, which used the same form factors
for charge and magnetic dipole interactions. For the typical direct
detection experiments, one has the recoiled energy from a few keV to a
hundred keV. Choosing a representative $E^{\rm ref}_R = 10$~keV, we
obtain the reference dark matter-nucleon scattering cross section
\beqa
\sigma^{\rm ref}_{\chi N} = \frac{(c_1-2 c_2)^2\,e^2 \,Z^2}{2\pi A^2} \, \frac{E^{\rm ref}_R m_p^2 (m_T + m_\chi)^2 } { m_T (m_p + m_\chi)^2 }  \approx 2 \times 10^{-49}~\mbox{cm}^2 \,,
\label{eq:majorana-DD-cross}
\eeqa
for $^{129}_{54}$Xe and the muon case with $m_\chi =50$~GeV,
$m_\phi=100$~GeV and $\lambda=1$. The current LUX results are not
sensitive to this cross section. We therefore do not show the direct
detection constraints on the Majorana fermion case in our plots. 

For the complex scalar case, the dominant contribution can be related to the charge radius operator
\beqa
{\cal L} \supset C\, \partial^\mu X \partial^\nu X^\dagger F_{\mu\nu} \,,
\eeqa
with the matched coefficient as $C(m_{e^i}, m_\psi)$ and the formula
\beqa
C(m_1, m_2) = \frac{\lambda^2 \, e}{16\pi^2}\left[ \frac{m_1^4 - 6 m_1^2 m_2^2 + m_2^4}{(m_1^2 - m_2^2)^3}  - \frac{4(m_1^2 + m_2^2) (m_1^4 - 5 m_1^2 m_2^2 + m_2^4)}{3 (m_1^2 - m_2^2)^4}\,\ln{\left(\frac{m_1}{m_2}\right)} 
\right] \,,
\eeqa
where $C(m_1, m_2) \propto (m_1 - m_2)$ in the limit of $m_1 - m_2 \ll 0$. In the limit of $m_1 \ll m_2$, one has
\beqa
C(m_1, m_2) = - \frac{\lambda^2 \, e}{16\pi^2\,m_2^2}\left[ 1 + \frac{2}{3} \ln{\left(\frac{m_1^2}{m_2^2}\right)} \right] \,.
\eeqa
The spin-independent dark matter-nucleus differential scattering cross section, at the leading order in $v^2$, is calculated to be
\beqa
\frac{d\sigma}{dE_R} = \frac{Z^2\,e^2\,C^2(m_{e^i}, m_\psi) m_T}{16\pi\,v^2} F_E^2(q^2) \,,
\eeqa
which has the same $v^2$ dependence as the ordinary spin-independent
scattering.  We obtain the total scattering cross section and the
averaged dark matter-nucleon cross sections
\beqa
\sigma_{XT} = \frac{Z^2\,e^2\,C^2(m_{e^i}, m_\psi)\,\mu^2_{XT}}{8\pi} \,,\qquad
\sigma_{XN} = \frac{Z^2\,e^2\,C^2(m_{e^i}, m_\psi)\,\mu^2_{XN}}{A^2\,8\pi} 
 \,.
\eeqa
The constraints on the model parameter space from LUX~\cite{Akerib:2013tjd} are shown in Fig.~\ref{fig:constraint-complex}.

\section{Dark Matter Indirect Detection}
\label{sec:indirect-detection}
The indirect detection of dark matter tries to observe the excess of
events in cosmic rays. If the dark matter annihilation cross section
is not $p$-wave suppressed, this is the most efficient way to test the
``WIMP miracle". For the three dark matter cases considered in this
paper, we only have the Dirac fermion case with a large indirect
detection signal.  We therefore work out the relevant predictions in
the Lepton Portal models for the Dirac fermion dark matter. We also
note that we have not considered the case of a degenerate spectrum with
co-annihilation.  Future indirect detection results from
CTA~\cite{Consortium:2010bc} could serve as the leading approach to
uncover this region of parameter space as emphasized in
Ref.~\cite{Garny:2012eb,Garny:2013ama}. 

The primary bounds on the Lepton Portal models from indirect detection
come from measurements of the high-energy positron flux.  Most
astrophysical processes generate more electrons than positrons, while
dark matter annihilations in the Lepton Portal model produce them in
equal amounts, leading to a distinctive excess in the positron
fraction, particularly at high energies for relatively heavy dark
matter.

Several experiments have measured the positron fraction at
high energies, but the cleanest measurement for the region of
interest was performed by the AMS-02 experiment.  They observed a rise
in the positron fraction above 10 GeV that cannot be conclusively
explained by known astrophysical sources~\cite{Aguilar:2013qda} (see
Ref.~\cite{Adriani:2008zr} for PAMELA results and
Ref.~\cite{FermiLAT:2011ab} for Fermi-LAT results).  The leading
candidate SM explanation for this excess at the time of this
publication is the generation and acceleration of positrons in
pulsars~\cite{Cholis:2013psa}. The possibility that this excess is due
to annihilations or decays of dark matter particles remains allowed.

Portions of Lepton Portal parameter space are excluded by the AMS-02
data regardless of the origin of the positron fraction rise, simply by
virtue of the fact that they produce a positron flux larger than
observed.  We determine the portion of parameter space excluded by
the AMS-02 results in this section.

We begin by calculating the differential flux of positrons and
electrons due to dark matter annihilation in the Lepton Portal model.
For a given annihilation cross-section, the flux is given 
by~\cite{Cirelli:2008id}
\begin{equation}
  \label{eq:electron-flux}
  \Phi_{e^\pm}(E) = B \frac{v_e}{4 \pi b(E)} \frac{1}{2} \left(\frac{\rho_{\odot}}{M_{\rm DM}}\right)^2 \int_E^{M_{\rm DM}} dE^\prime f_{{\rm inj},e^\pm} (E^\prime)\, I\left[\lambda_D(E, E^\prime)\right]\,.
\end{equation}
The particle physics inputs to this calculation are encoded entirely
in the dark matter mass, $M_{\rm DM}$, the injection spectrum, $f_{\rm
  inj}$, and the electron/positron velocity, $v_e \approx c$.  The
dark matter injection spectrum is given by
\begin{equation}
  \label{eq:2}
  f_{{\rm inj},e^\pm}(E) = \sum_k \langle \sigma v \rangle_k \frac{dN_{e^\pm}^k}{dE} \,,
\end{equation}
where the sum is over processes with an electron/positron in the final
state, $\langle \sigma v \rangle_k$ is the thermally averaged
cross-section for annihilation via process $k$, and $dN_{e^\pm}^k/dE$
is the expected number of electrons/positrons with energy between $E$
and $E + dE$ produced by the annihilation.  For the case ${\rm DM} +
{\rm DM} \to e^+ e^-$, using the fact that the annihilations occur
between non-relativistic DM particles, we find
\begin{equation}
  \label{eq:3}
  \frac{dN_{e^+}^k}{dE} = \frac{dN_{e^-}^k}{dE} = \delta(E - M_{\rm DM})\,.
\end{equation}
The muon and tau cases have been studied in Ref.~\cite{Bai:2009ka}.  For the muon case,
\begin{equation}
  \label{eq:9}
  \frac{dN_{e^+}^k}{dE} = \frac{dN_{e^-}^k}{dE} = \frac{1}{3 M_{\rm DM}} (5 - 9 x^2 + 4 x^3)\times \theta(M_{\rm DM} - E)  \,,
\end{equation}
where $x = E/M_{\rm DM}$ and $\theta$ is the Heaviside theta function.  For the tau case, the spectrum is generated using \texttt{Pythia} \cite{Sjostrand:2007gs} and is fitted by~\cite{Bai:2009ka} 
\begin{multline}
  \label{eq:9}
  \frac{dN_{e^+}^k}{dE} = \frac{dN_{e^-}^k}{dE} = \frac{1}{M_{\rm DM}} (e^{-ˆ'97.716 x^5 + 223.389 x^4 - 193.748 x^3 + 82.595 x^2 - 22.942 x + 2.783} \\+ e^{-15.575x^3 + 15.79 x^2 - 18.083 x+ 0.951}) \theta(M_{\rm DM} - E) \,.
\end{multline}

The remaining factors in Eq.~(\ref{eq:electron-flux}) are purely
astrophysical.  $B$, taken to be 1, is a boost factor that accounts
for possible local clumping of dark matter. The energy loss
coefficient $b(E) = E^2/(\mbox{GeV} \cdot \tau_E)$ with
$\tau_E=10^{16}$~s, is defined by the diffusion equation
\begin{equation}
  \label{eq:5}
  \frac{\partial f}{\partial t} - K(E) \nabla^2 f - \frac{\partial}{\partial E} \left[b(E) f \right] = Q \,,
\end{equation}
where $K(E)$ is the diffusion coefficient and $Q$ is the annihilation
injection term.  $\rho_{\odot}$ is the local dark matter density.  $I$
is the ``halo function'', depending on the diffusion length
$\lambda_D$.  All of these quantities are described and fit to
functions in \cite{Delahaye:2007fr,Cirelli:2008id} for a variety of assumptions ranging from
conservative to optimistic.  In this study, we use the flux as
determined by the min, med, and max set of assumptions from
\cite{Delahaye:2007fr} to represent minimal, medium, and maximal fluxes
attainable by varying the astrophysical assumptions.

\begin{figure}[t!b]
  \centering
  \begin{tabular}{c c c}
  \hspace{-3mm}
    \includegraphics[width=0.33\textwidth]{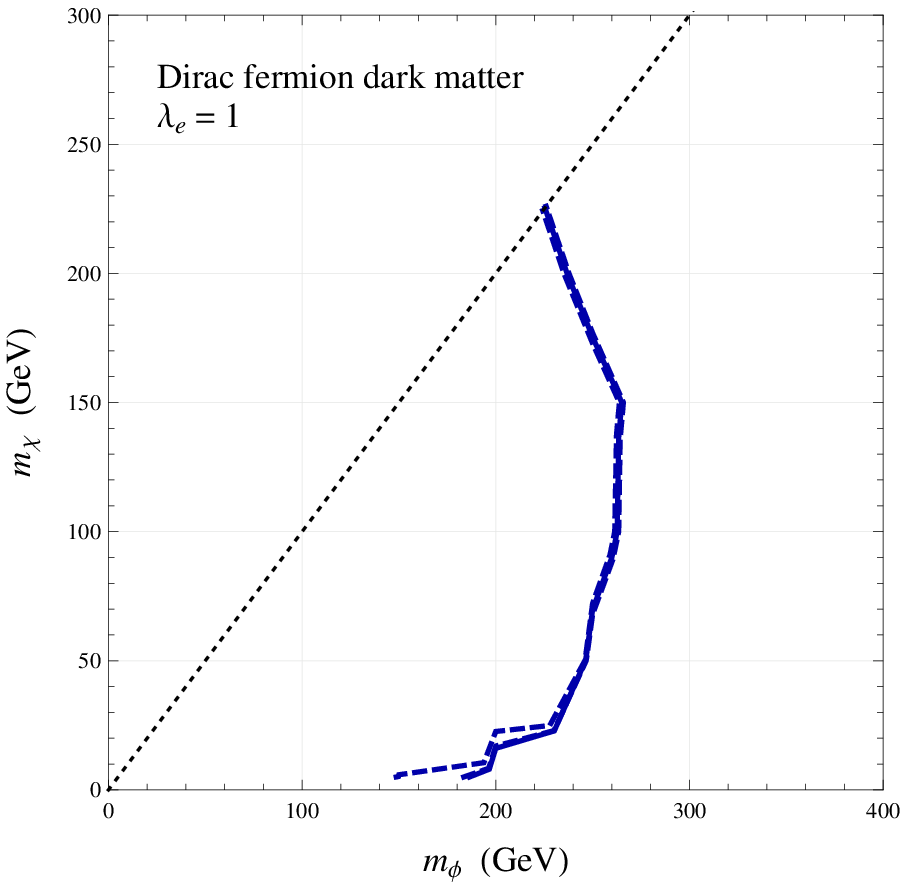} &   \hspace{-3mm}
    \includegraphics[width=0.33\textwidth]{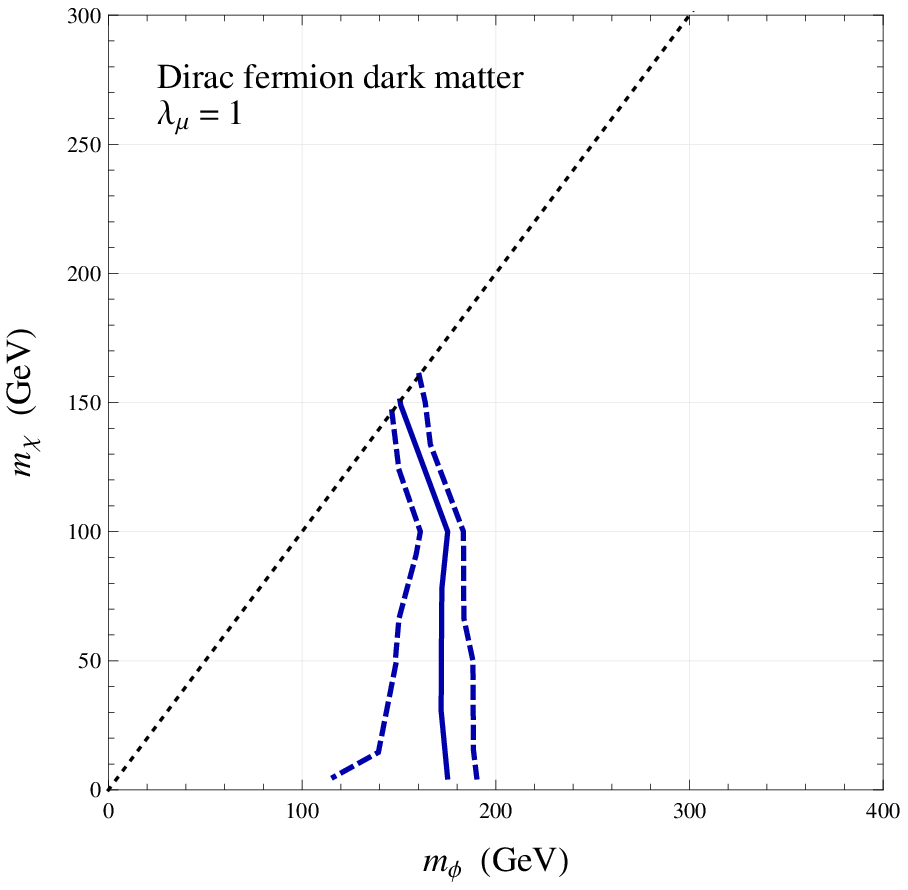} &   \hspace{-3mm}
     \includegraphics[width=0.33\textwidth]{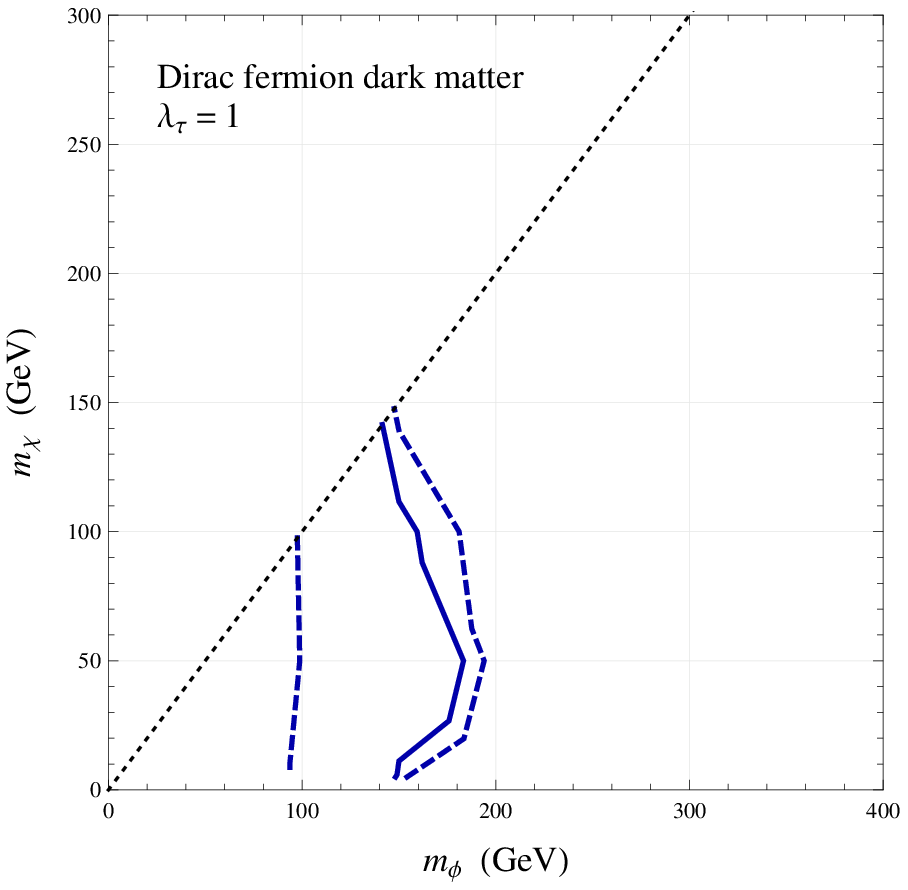}  \\
   {\small (a) } & {\small (b)}   & {\small (c) } 
     \end{tabular}
  \caption{Indirect detection constraints on the Lepton Portal model
    for the Dirac fermion dark matter with coupling to (a) electrons; (b) muons;  (c) taus. The left and right  outer dashed lines represent the ``minimal'' and ``maximal'' astrophysical assumptions, respectively.  The middle and solid line represents the ``medium'' astrophysical assumption.}
  \label{fig:indirect}
\end{figure}

We now determine the excluded regions of Lepton Portal parameter space
assuming that the observed flux is enitrely due to SM processes.  To
be conservative regarding astrophysical positron sources, we determine
that a model is excluded if it predicts 
a \emph{total} positron flux more than $2\sigma$ in excess of that
measured by AMS-02 in any energy bin (see
Refs.~\cite{Bergstrom:2013jra,Ibarra:2013zia} for model-independent
constraints).  The total number of positrons predicted by AMS-02 is
given by the product of the fraction spectrum~\cite{Aguilar:2013qda}
and the $e^-+e^+$ spectrum~\cite{AMS-electron-positron} 
\begin{equation}
  \label{eq:6}
  \frac{d\Phi_{e^+}}{dE}|_{\rm AMS} = f_{e^+,{\rm AMS}}(E) \, \frac{d\Phi_{e^- + e^+}}{dE}\arrowvert_{\rm AMS} \,.
\end{equation}

Majorana fermion and complex scalar dark matter cases have a
velocity-suppressed annihilation cross-section, ensuring that the
indirect detection signal is too small to be observed.  For Dirac
fermion dark matter, non-zero $s$-wave annihilation leads to
constraints from AMS-02. The formula for the annihilation
cross-section in the non-relativistic limit is given by
Eq.~(\ref{eq:dirac-ann}) for a Dirac Fermion by neglecting the
$p$-wave parts. As a benchmark, we also take the 
coupling $\lambda = 1$. The resulting constraints under the
conservative set of assumptions are shown in
Fig.~\ref{fig:indirect}. One can see that our conservative constraints
require the mediator masses to be above 100-300 GeV for different
flavor and propagation model assumptions. For the electron coupling
case, the limits for the three different propagation models are
similar to each other. This is because the electron/positron
propagation difference decreases at an energy close to the dark matter
mass and the constraints from AMS-02 mainly come from high energy
bins.

\section{Collider Constraints and Searches}
\label{sec:collider}
\begin{figure}[th!]
\begin{center}
\hspace*{-0.75cm}
\includegraphics[width=0.48\textwidth]{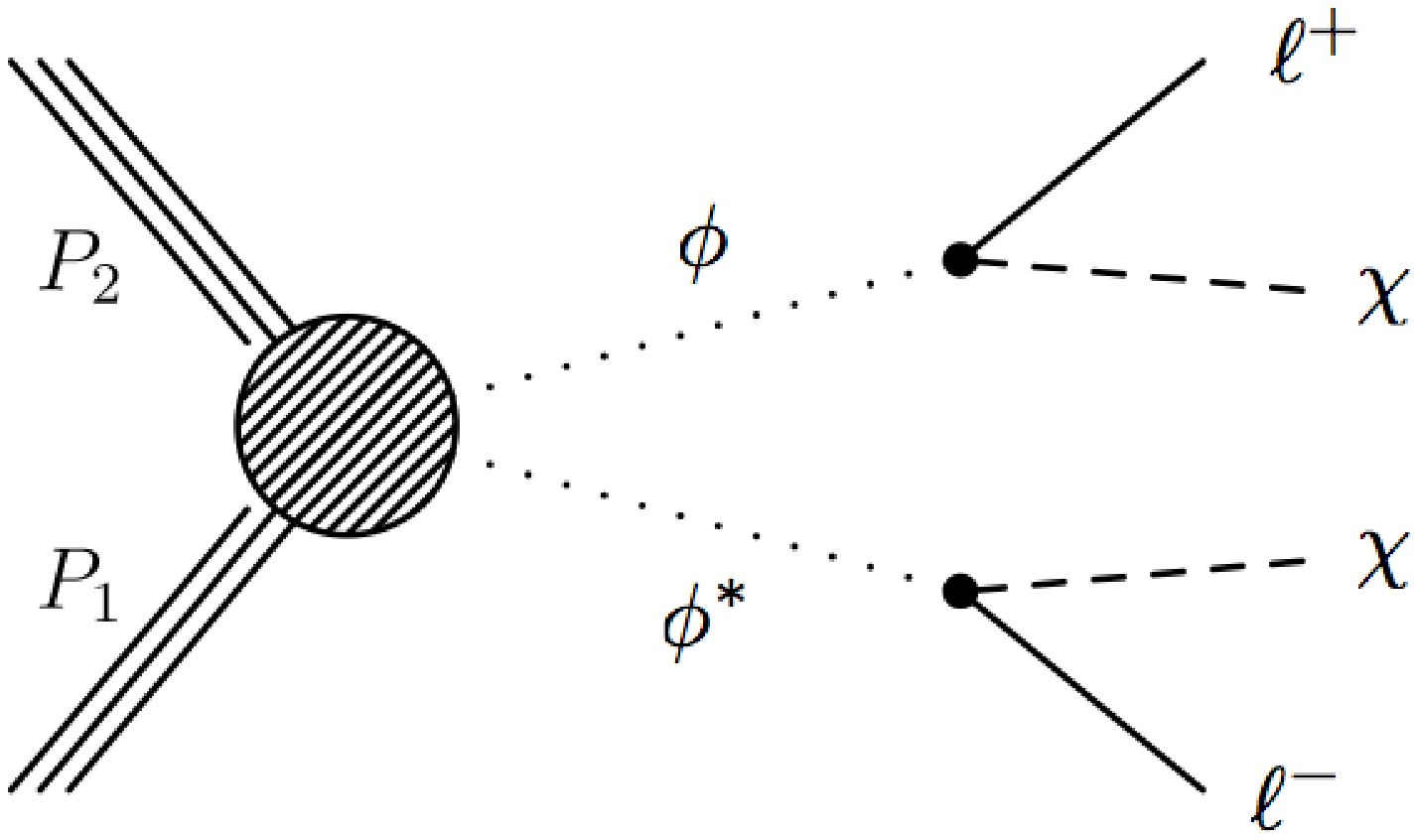}
\includegraphics[width=0.48\textwidth]{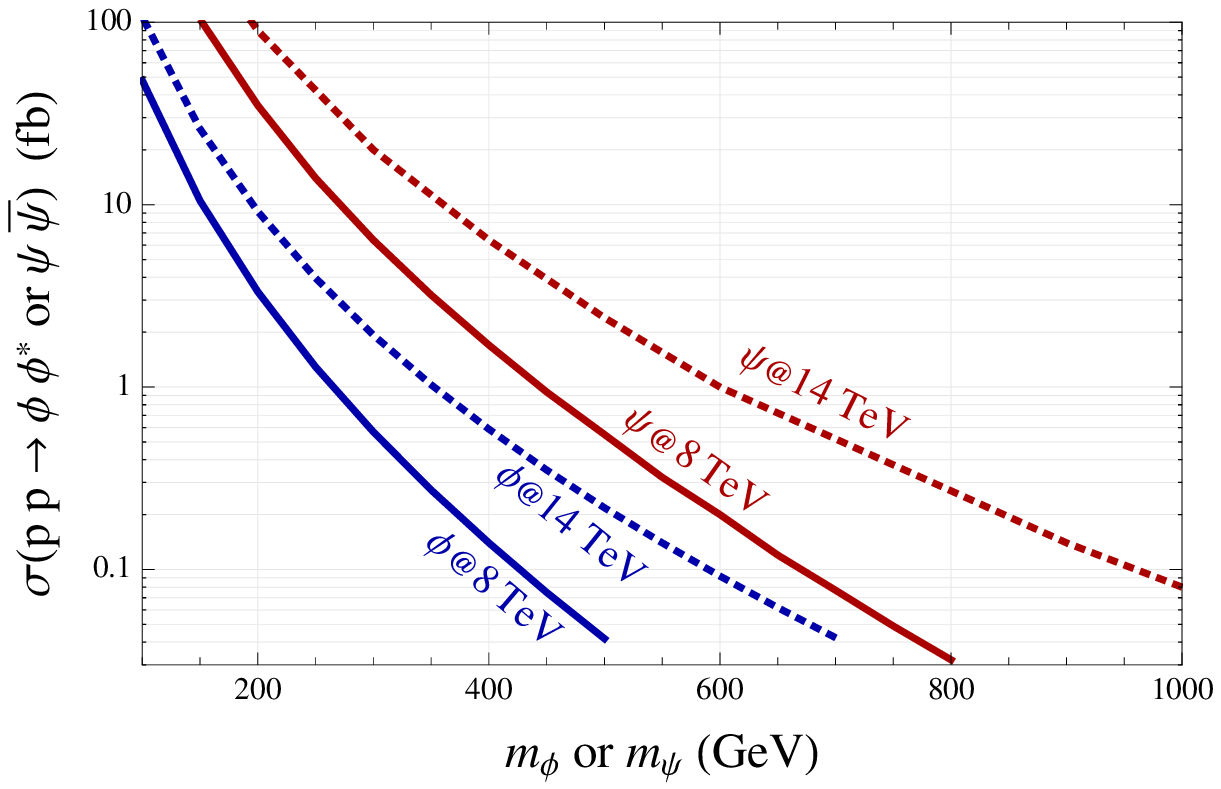}
\caption{Left panel: Feynman diagram for the complex scalar mediator production and decay in the fermion dark matter models. Right panel: the production cross sections for the complex scalar and vector-like fermion mediators at the LHC.}
\label{fig:feyn-diagram}
\end{center}
\end{figure}

At hadron colliders, the signature of Lepton Portal models comes from pair
productions of the mediator via the Drell-Yan process. The produced
mediator particles then decay into the dark matter particles plus
leptons. The signature at hadron colliders is thus same-flavor, opposite-sign
dilepton plus missing transverse energy, which is also the standard
signature for searching for sleptons in the MSSM at colliders. We show the
production and decay processes in the left panel of
Fig.~\ref{fig:feyn-diagram} for a complex scalar mediator.
In the right-panel of Fig.~\ref{fig:feyn-diagram}, we show the
production cross sections of mediators, $\phi$ and $\psi$, for
different masses at the LHC with both 8 TeV and 14 TeV center of mass
energy. The $\phi +\phi^*$ production cross section is the same as a
single-flavor right-handed slepton in
MSSM~\cite{Beenakker:1996ed,Fuks:2013lya}. In the complex scalar dark
matter case, the fermion mediator can be thought as a vector-like
fermion with the same electroweak quantum number as the right-handed
electron. Its production cross sections are much larger than the
scalar mediator one with the same mass. We will show later that the
discovery sensitivity for this case is much better than the scalar
mediator case.

Both ATLAS and CMS colaborations have searches for new physics in the
$\ell^+\ell^-+\mbox{MET}$ channel. The latest results from ATLAS with
20.3 fb$^{-1}$ at 8 TeV have constrained the selectron and smuon
masses to be above around 240~GeV~\cite{ATLAS-CONF-2013-049} for a
light neutralino mass by summing the signal events from both selectron
and smuon. For the Lepton Portal model with coupling only to a
single flavor lepton, the signal production cross section is reduced
by a factor of two. As a result, the constraint on the mediator mass
is weaker and is around 170~GeV. A similar result has been obtained by
the CMS collaboration~\cite{CMS-PAS-SUS-13-006}, although different
kinematic variables were used. The CMS collaboration has used $M_{{\rm
    CT}\perp}$~\cite{Matchev:2009ad}, which is related to the
contransverse mass $M_{\rm CT}$~\cite{Tovey:2008ui} (see also
Ref.~\cite{Buckley:2013kua} for the super-razor variable). On the
other hand, the ATLAS collaboration has used the
$M_{T2}$~\cite{Lester:1999tx,Barr:2003rg,Cheng:2008hk,Konar:2009qr}
variable to reduce the SM backgrounds (see also
Refs.~\cite{Kats:2011qh,Bai:2012gs,Kilic:2012kw,Bai:2013ema} for
recent applications on searching for stops).  In our analysis, we
concentrate on following the analysis of the ATLAS collaboration
and use the $M_{T2}$ variable to explore the discovery and exclusion
sensitivities at both 8 TeV and 14 TeV LHC.  

\begin{figure}[th!]
\begin{center}
\hspace*{-0.75cm}
\includegraphics[width=0.48\textwidth]{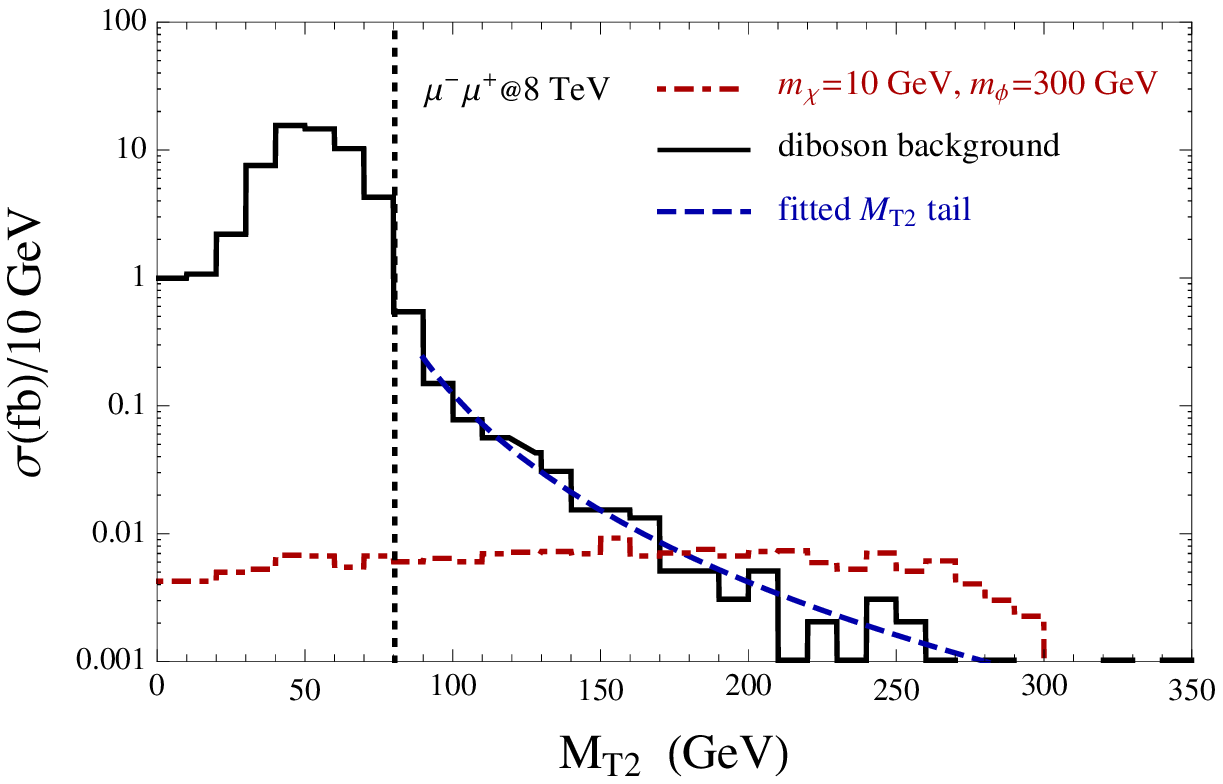} \hspace{0.3cm}
\includegraphics[width=0.48\textwidth]{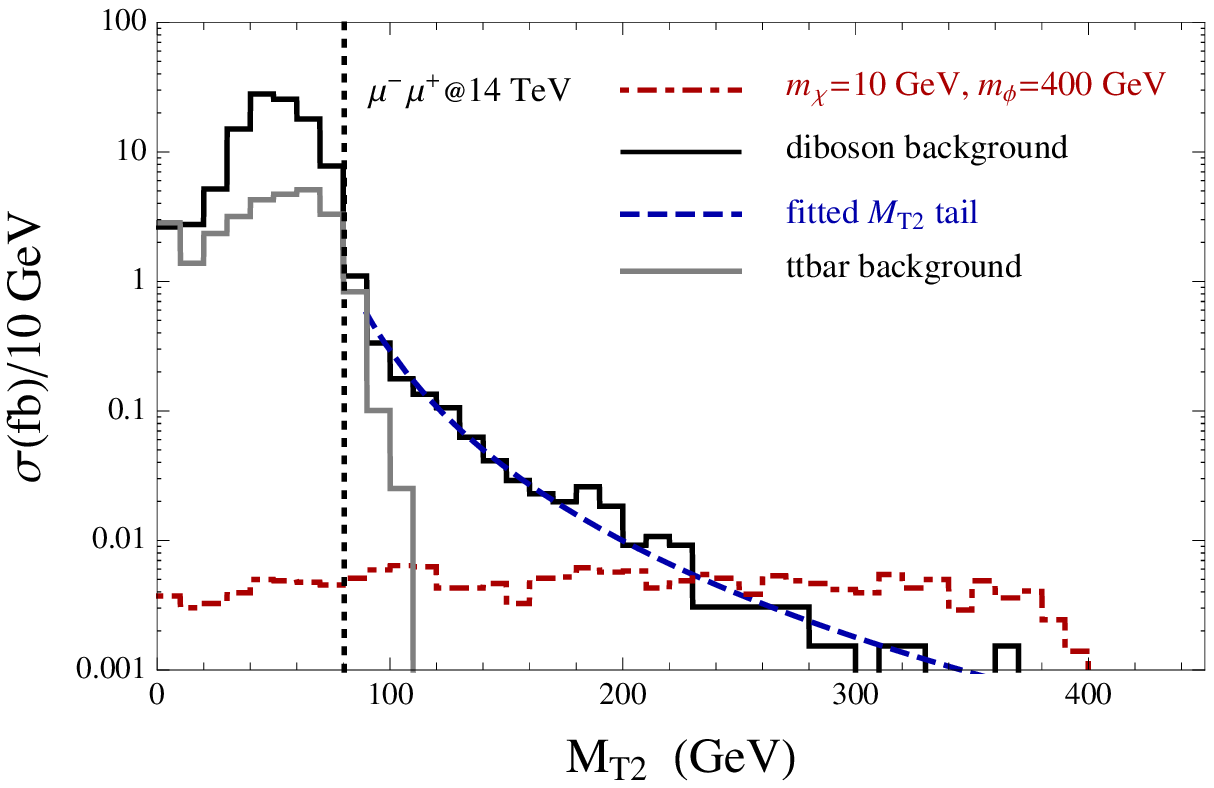}
\caption{Left panel: the dilepton $M_{T2}$ distributions for the diboson background and the signal events. The missing particle mass is assumed to be zero. The blue and dashed line is from the fitted function in Eq.~(\ref{eq:fit-function}) with $\eta=2.0$. The vertical and dotted line indicates the reference $W$ gauge boson mass. Right panel: the same as the left one but for the 14 TeV LHC together with the $t\bar t$ background. The same value $\eta=2.0$ is used for the fit function of Eq.~(\ref{eq:fit-function}).}
\label{fig:MT2-dist}
\end{center}
\end{figure}

Other than the basic cuts on selecting the objects, the ATLAS searches
have required two leptons with opposite signs and either the same or
different flavors. They also veto events with a jet above 20 GeV,
events with $|m_{\ell\ell} -m_Z| < 10$~GeV and events with $M_{T2} <
90 (110)$~GeV. After those cuts, the main backgrounds are from diboson
productions. The dilepton $M_{T2}$ variable will be the most sensitive
one for searching for higher mediator masses at the 14 TeV. It is
defined as  
\beqa
M_{T2} = \mbox{min}\left\{   \bigcup_{ \vec{p}^T_1 + \vec{p}^T_2 = \vec{E}_T^{\rm miss} } \mbox{max} {\Big[} M_T(\vec{p}_{\ell_1},  \vec{p}^T_1), M_T(\vec{p}_{\ell_2},  \vec{p}^T_2) {\Big]}  \right\}  \, ,
\label{eq:mt2}
\eeqa
with the transverse mass in terms of the lepton momentum
$\vec{p}_{\ell_i}$ and the guessed missing particle (massless)
transverse momentum $\vec{p}^T_{i}$.  As we know from the discovery of
the $W$ gauge boson, the transverse mass of the electron and neutrino
is bounded from above by the $W$ gauge boson
mass~\cite{Arnison:1983rp,vanNeerven:1982mz,Barger:1983wf,Smith:1983aa}. Imposing
a cut on $M_{T2}$ to be above the $W$ gauge boson mass can therefore
dramatically reduce the dominant diboson backgrounds. The tail of the
dilepton $M_{T2}$ becomes the leading background, especially for a
heavy mediator mass, as can be seen in Fig.~\ref{fig:MT2-dist}. To
estimate the current bounds on this model, we calculate LO
cross-sections for the full process using \texttt{MadGraph}
\cite{Alwall:2011uj} using a model constructed by \texttt{FeynRules}
\cite{Christensen:2008py}. The events are showered and hadronized
using \texttt{Pythia} \cite{Sjostrand:2007gs}, then the hadrons are
clustered into jets using \texttt{PGS} \cite{PGS}.

Motivated by the method of measuring the $W$ gauge boson width using
the transverse tail distribution~\cite{Smith:1983aa,Abazov:2009vs}, we
suspect that the tail of $M_{T2}$ should be generated from off-shell
$W$ gauge bosons and could follow the general Breit-Wigner
distribution.  We introduce the following parametrical function to fit the tail distribution
\beqa
F\left(M_{T2}\right) = \frac{N_0}{ \left[ \eta M_{T2}^2 - M_W^2 \right]^2 +  \eta^2 M_{T2}^4 \,\Gamma_W^2/M_W^2 }  \,.
\label{eq:fit-function}
\eeqa
Here, $N_0$ is the overall normalization and $\eta > 1$ is suggested
by the fact that the invariant mass of the $W$ gauge boson propagator
is above the corresponding transverse mass. In Fig. 7, one can see
that this Breit-Wigner distribution fits the tail pretty well. With a
better understanding of the main background, the discovery reach of
Lepton Portal dark matter can be extended.

\begin{figure}[th!]
\begin{center}
\hspace*{-0.75cm}
\includegraphics[width=0.45\textwidth]{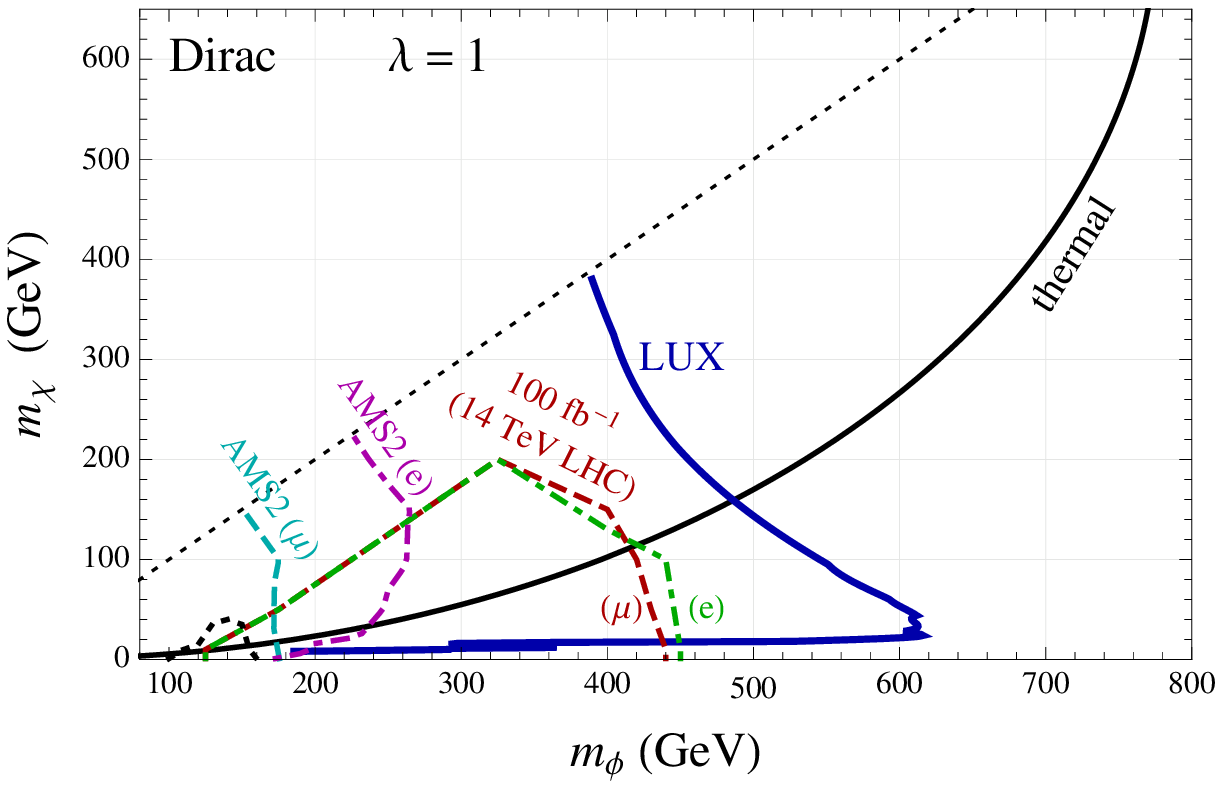} \hspace{3mm}
\includegraphics[width=0.45\textwidth]{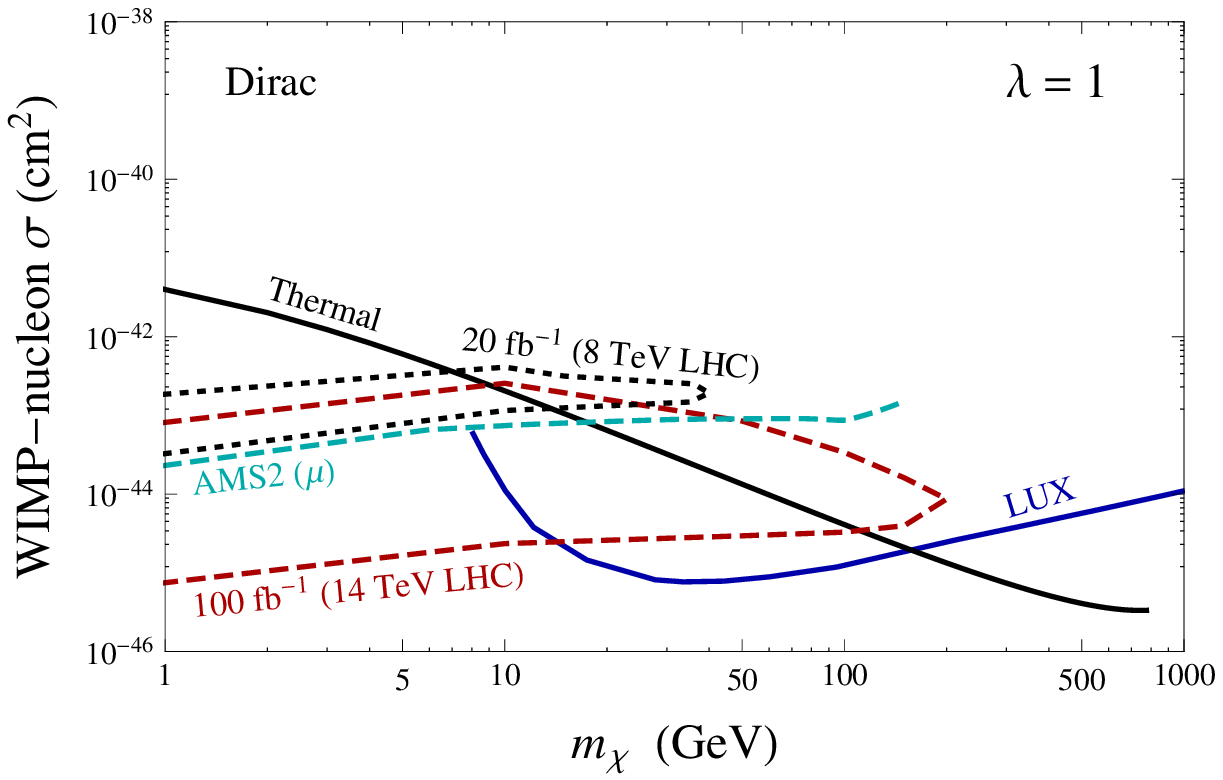}
\caption{Left panel: the constraints on the dark matter and its mediator masses for the Dirac fermion case. The dotted and black line is the current constraint on the muon case from the 8 TeV LHC with 20 fb$^{-1}$~\cite{ATLAS-CONF-2013-049}. Right panel: the dark matter-nucleon scattering cross section as a function of dark matter mass from different searches.}
\label{fig:constraint-dirac}
\end{center}
\end{figure}
\begin{figure}[th!]
\begin{center}
\hspace*{-0.75cm}
\includegraphics[width=0.45\textwidth]{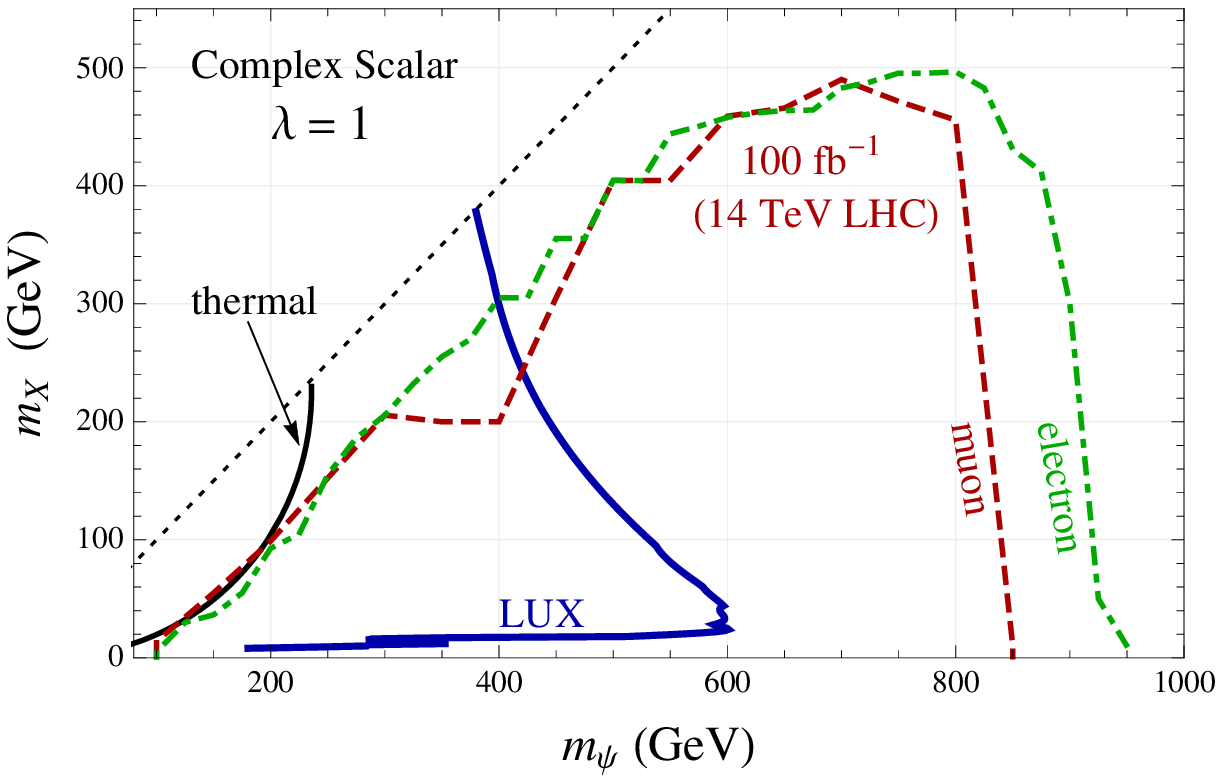} \hspace{3mm}
\includegraphics[width=0.45\textwidth]{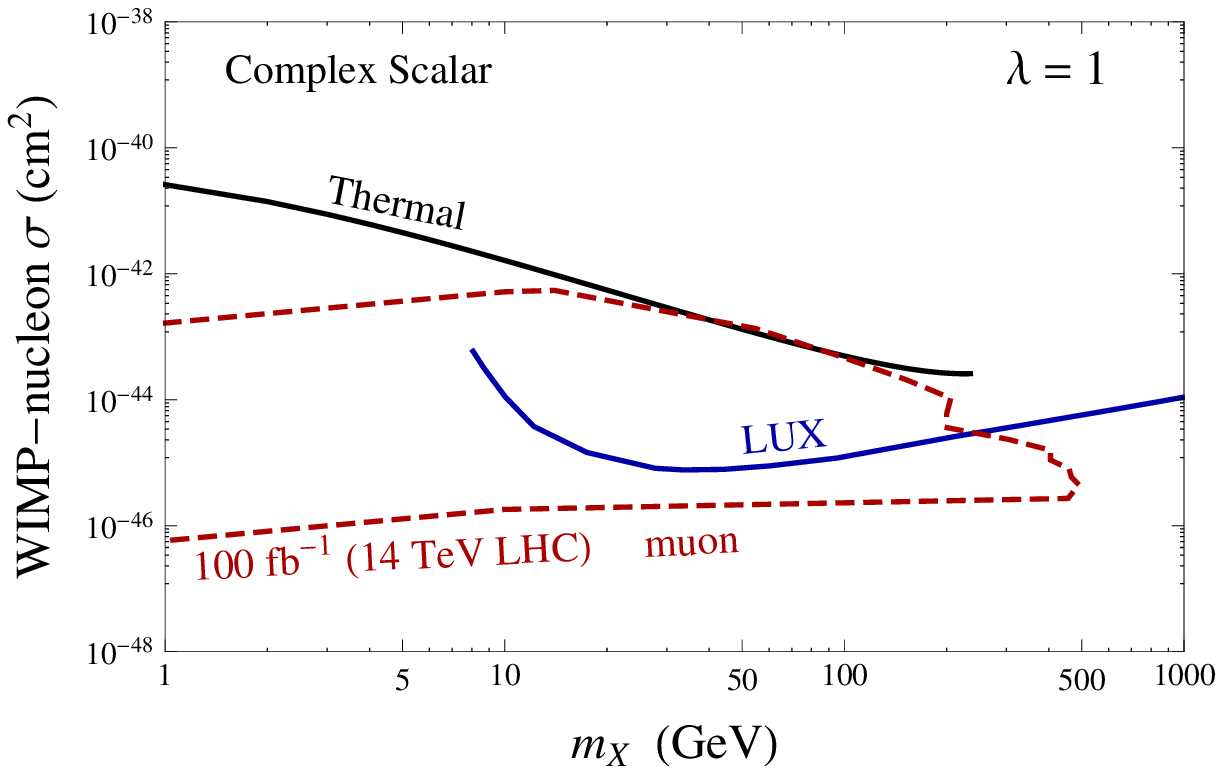}
\caption{The same as Fig.~\ref{fig:constraint-dirac} but for the complex scalar case. Because of the $p$-wave suppression of the dark matter annihilation cross section, the indirect detection constraints become very weak and are not shown here.}
\label{fig:constraint-complex}
\end{center}
\end{figure}

We simulate the signal and background events at the 14 TeV LHC and
work out the 90\% CL exclusion region on the model parameter space for
100 fb$^{-1}$ luminosity in the left panel of
Fig.~\ref{fig:constraint-dirac} and Fig.~\ref{fig:constraint-complex}
for fixed $\lambda=1$. Specifically, for a given mass point, we have
calculated the sensitivities for three different cuts: $M_{T2} \ge
100, 200, 300$~GeV and chosen the most sensitive one as the potential
reach. We also translate the LHC reach into the potential constraints
on the dark matter-nucleon scattering cross section in the right panel
of Fig.~\ref{fig:constraint-dirac} and
Fig.~\ref{fig:constraint-complex}. For both Dirac fermion and complex
scalar dark matter cases, the LHC searches have a better sensitivity
for a light dark matter with a mass below 10 GeV. For the complex
scalar dark matter case, the LHC has a better reach than direct
detection experiments with dark matter masses up to around
500~GeV. This is due to the large production cross sections of
vector-like fermion mediators at the LHC. The LHC reaches for the
electron and muon cases are not different significantly from each
other. The electron case has a larger acceptance and hence a better
limit.

The collider constraints for the Majorana fermion dark matter case are
identical to the Dirac fermion case, since the mediator production
cross section determines the sensitivity. As discussed in
Eq.~(\ref{eq:majorana-DD-cross}), the direct detection cross section
is very small for the Majorana fermion case. The indirect detection is
$p$-wave suppressed or suppressed by ${\cal O}(v^2/c^2 \approx
10^{-6})$.  The collider search is the most relevant one and
can probe a large region of unexplored parameter space.

\section{Discussion and Conclusions}
\label{sec:conclusion}
We want to first emphasize the importance of colliders for discovering
or excluding the Lepton Portal dark matter.  The signature with the
same-flavor and opposite-sign dilepton plus missing energy is a pretty
clean one. The $M_{T2}$ cut can be imposed to make almost background
free. As a result, the discovery reach is purely determined by the
signal production cross section times the acceptance. For a large mass
splitting between dark matter and its partner, the signal acceptance
is large, so the discovery reach is limited by the signal cross
section. From Fig.~\ref{fig:feyn-diagram}, one can see a large
increase of the mediator production cross sections from 8 TeV to 14
TeV and a discovery of dark matter signals at the LHC may happen in
the near future.

In our analysis, we have considered both the electron and muon cases
and neglected the tau lepton case. We anticipate a slightly weaker
limit from the LHC because of the tau-tagging and mis-tagging
efficiencies.  Another parameter region that we have ignored is the
co-annihilation region.  The collider searches become less sensitive
because the leptons from the mediator decays are either too soft to
pass the basic cuts or generate insufficient $M_{T2}$ and would be buried in
the SM backgrounds.  In the extremely degenerate region, one could
include an additional jet, photon, $W$ and $Z$ gauge bosons from initial
state radiation to gain sensitivity.

In summary, we have studied Lepton Portal dark matter for three
cases: Majorana fermion, Dirac fermion and complex scalar dark
matter. For direct detection, the majorana fermion case has a very
small predicted event rate because of the leading operator of the dark
matter coupling to photon has an additional velocity suppression. On
the other hand, the direct detection signals for the Dirac fermion and
complex scalar cases are not suppressed. In terms of indirect detection, since
only the Dirac fermion case has non-zero $s$-wave annihilation,
AMS-02 has the best coverage for its model parameter space. At
colliders, the LHC has better reaches for the light dark matter mass
region than the direct detection experiments.  For the complex scalar
case, the 14 TeV LHC with 100 fb$^{-1}$ can cover mediator masses up
to 800 GeV and provides a constraint on spin-independent dark
matter-nucleon scattering cross section as low as
$2\times10^{-46}$~cm$^2$ for dark matter masses up to 500 GeV and a
unit coupling. 

\subsection*{Acknowledgments} 
We thank Matthew Buckley, Zackaria Chacko, Spencer Chang and
especially Wai-Yee Keung for useful discussions and
comments. J.~Berger would like to thank the Aspen Center for
Theoretical Physics for their hospitality during the early stages of
this work.  Y.~Bai is supported by the U. S. Department of Energy under the contract
DE-FG-02-95ER40896. SLAC is operated by Stanford University for the US
Department of Energy under contract DE-AC02-76SF00515. 

\begin{appendix}
\section{Non-relativistic Correspondence of Photon Couplings}
\label{app:a}
At $v = 0$, it is well known that there are only four ways in which a
particle with spin $\vec{S}$ can couple to the electromagnetic field:
charge operator $\Phi$, electric dipole moment $e\,\vec{S}\cdot \vec{E}$,
magnetic dipole moment $e\,\vec{S} \cdot \vec{B}$, and anapole moment
$e\,\vec{S}\cdot (\nabla \times \vec{B})$\footnote{In principle, there could also be magnetic
monopoles, in concert with ``electric'' anapoles.  Magnetic monopoles
violate $P$ and $T$.  Electric anapoles violate $C$ and $T$.  Without
magnetic monopoles, there is no operator that violates $C$ and $T$.}.  The properties of these
operators under $C$, $P$ and $T$ are shown for future reference in Table
\ref{tab:cpt-nr}.

\begin{table}[hb!]
\renewcommand{\arraystretch}{1.3}
  \centering
  \begin{tabular}{c|c c c}
    \hline \hline
    Operator & $C$ & $P$ & $T$ \\
    \hline
    $e\,\Phi$ & $+$ & $+$ & $+$ \\
    $e\,\vec{S} \cdot \vec{E}$ & $+$ & $-$ & $-$ \\
    $e\,\vec{S} \cdot \vec{B}$ & $+$ & $+$ & $+$ \\
    $e\,\vec{S} \cdot (\nabla \times \vec{B})$ & $-$ & $-$ & $+$ \\
    \hline \hline
  \end{tabular}
  \caption{$C$, $P$, $T$ properties of the non-relativistic couplings to photons.}
  \label{tab:cpt-nr}
\end{table}

In the non-relativistic limit, any operator coupling $\bar{\chi}$,
$\chi$, and $A^\mu$ should reduce to one or more of the above forms,
up to corrections of $\mathcal{O}([\nabla^2]^i)$ (radius corrections)
and $\mathcal{O}([\vec{v}]^i)$ which are fixed by Lorentz invariance.
Based on the $C$, $P$ and $T$ properties of a given operator, one can determine
which operator contributes.  For any non-renormalizable operator,
there cannot be a direct correspondence to $\Phi$, since gauge
invariance demands dependence on $\vec{E}$ and $\vec{B}$ only.  There
may, however, still be charge radius terms from $\nabla\cdot\vec{E} = \nabla^2 \Phi$.

Coming back to the operators ${\cal O}_1$ and ${\cal O}_2$ from
Eq.~\eqref{eq:dirac-operator}, we further break these operators up to
highlight their contributions from operators with different $C$, $P$
and $T$ properties.  We define
\begin{eqnarray}
  \label{eq:cpt-operators}
  e\,{\cal O}_1^V & = & e\, (\bar{\chi} \gamma^\mu \partial^\nu \chi + {\rm h.c.})
  F_{\mu\nu} \,, \nonumber\\
  e\,{\cal O}_1^A & = & - e\, (\bar{\chi} \gamma^\mu \gamma^5 \partial^\nu \chi + {\rm h.c.}) F_{\mu\nu}  \,, \nonumber\\
  e\,{\cal O}_2^V & = & e\, \epsilon_{\mu\nu\alpha\beta} (i \bar{\chi} \gamma^\mu \partial^\nu \chi + {\rm h.c.}) F^{\alpha\beta} \,, \nonumber\\
  e \,{\cal O}_2^A & = & - e\, \epsilon_{\mu\nu\alpha\beta} (i \bar{\chi} \gamma^\mu \gamma^5 \partial^\nu \chi + {\rm h.c.}) F^{\alpha\beta}  \,.
\end{eqnarray}
Then operators ${\cal O}_1^V$ and ${\cal O}_2^A$ have the same $C$, $P$ and $T$ properties
of a charge or a magnetic dipole operator, while ${\cal O}_1^A$ and ${\cal O}_2^V$
have the properties of an anapole.  ${\cal O}_1^V$ can easily be rewritten
using integration by parts as $\bar{\chi} \gamma^\mu \chi \partial_\nu
F^{\mu\nu} = J^\mu_\chi\, \partial_\nu F^{\mu\nu}$, making the correspondence
to the charge radius operator evident.  ${\cal O}_2^A$, on the other hand,
contains the axial current which is proportional to spin in the
non-relativistic limit; it maps onto a magnetic dipole operator.  To
order $v^0$, we then have
\begin{eqnarray}
  \label{eq:cpt-expansion}
  e\, {\cal O}_1^V & \approx & e\, \xi^\dagger \xi \,\nabla \cdot \vec{E} \,, \nonumber\\
  e \,{\cal O}_1^A & \approx & - e\, \xi^\dagger \vec{S} \xi \cdot (\nabla \times \vec{B})\,, \nonumber\\
  e \,{\cal O}_2^V & \approx & 2 e\, \xi^\dagger \vec{S} \xi \cdot (\nabla \times
  \vec{B})\,, \nonumber\\
  e \,{\cal O}_2^A & \approx & - 4 m\, e\, \xi^\dagger \vec{S} \xi \cdot \vec{B},
\end{eqnarray}
where $\xi$ is the fermion wave-function, such that $e\, \bar{\xi}
\xi$ is the charge density and $e \,\xi^\dagger \vec{S} \xi$ is the
magnetization.

It may be further worth noting that these operators are not all
independent.  We have the following exact identities:
\begin{equation}
  \label{eq:4}
  2 {\cal O}_1^A + {\cal O}_2^V = 0\,,\qquad 2 {\cal O}_1^V + {\cal O}_2^A = 2 m\, {\cal O}_{\rm dipole}\,,
\end{equation}
where ${\cal O}_{\rm dipole} = \bar{\chi} \sigma^{\mu\nu} \chi F_{\mu\nu}$ is the
dimension-five anomalous magnetic moment operator.  These are obtained
using gamma matrix identities and the Dirac equation.

\section{Lepton $g-2$}
\label{sec:g-2}
In this appendix, we consider additional
contributions to the lepton anomalous magnetic moments, which could
also set constraints on our model parameters.  Among the different flavors,
the one that provides the most stringent constraint is the
$a_\mu=(g-2)_\mu/2$.  On the other hand, there is a disagreement of
more than 3$\sigma$ between the theoretical prediction and the
experimental measurement on this quantity.  The latest analysis of the
hadronic contributions gives an SM prediction
of~\cite{Hagiwara:2011af}
\beqa
a^{\rm SM}_\mu = (11659182.8 \pm 4.9)\times 10^{-10} \,,
\eeqa
while the experimental measured value is higher and is~\cite{Bennett:2006fi, Roberts:2010cj} 
\beqa
a^{\rm EXP}_\mu =  (11659208.9 \pm 6.3)\times 10^{-10}\,.
\eeqa
The difference is
\beqa
a^{\rm EXP}_\mu  - a^{\rm SM}_\mu = (26.1\pm 8.0)\times 10^{-10}\,,
\label{eq:g-2-diff}
\eeqa
which corresponds to a 3.3$\sigma$ discrepancy (see
Ref.~\cite{Blum:2013xva} for a recent review and lattice QCD
calculations for the SM prediction).

\begin{figure}[th!]
\begin{center}
\hspace*{-0.75cm}
\includegraphics[width=0.45\textwidth]{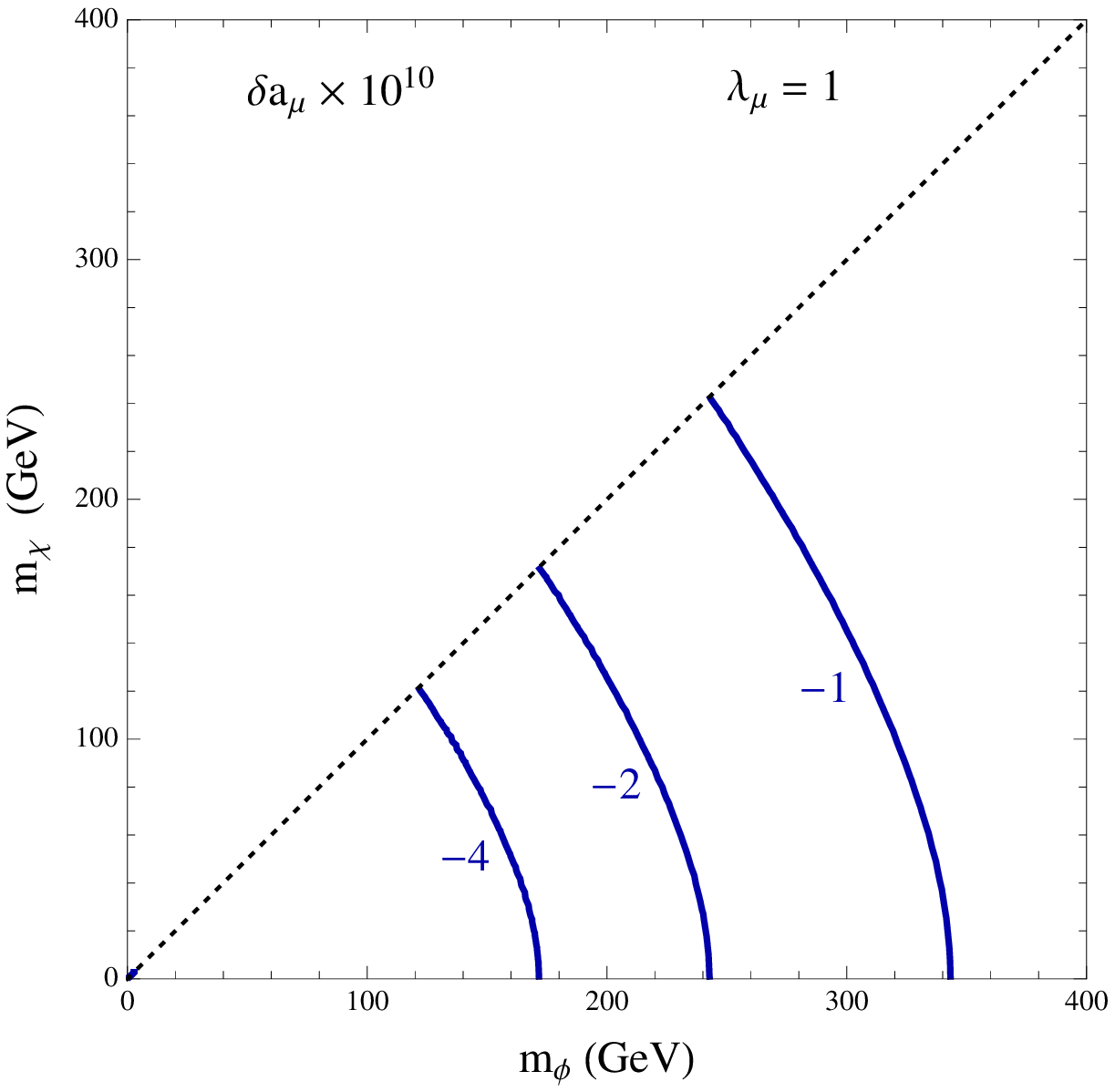} \hspace{3mm}
\includegraphics[width=0.45\textwidth]{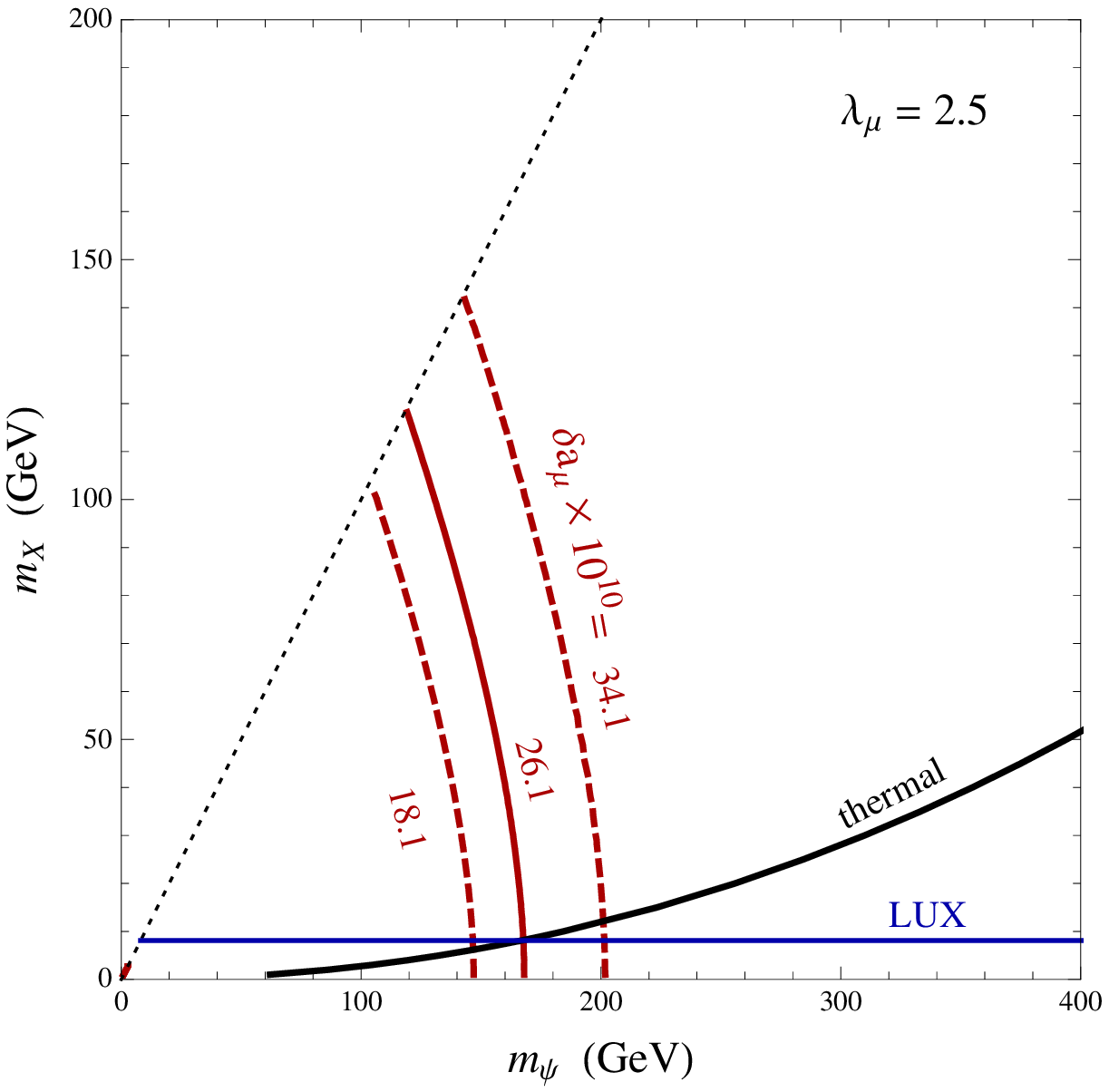}
\caption{Left panel: the contours of dark matter contributions to
  $\delta a_\mu$ in the fermionic dark matter case. Right panel: for the complex scalar dark matter model, the
  solid line indicates parameter space giving a contribution to
  $(g-2)_\mu$ equal to the central value of the discrepancy.  The two dashed lines are the one
  sigma boundaries from Eq.~(\ref{eq:g-2-diff}). The region above the blue line is excluded by the direct detection results from LUX.}
\label{fig:mu-g-2}
\end{center}
\end{figure}

The lepton-portal dark matter could explain such a discrepancy. We
check both parameter space that can fit the data and are allowed by
the $a_\mu$ data. For Majorana (also for Dirac) fermion dark matter,
the calculation has been done in the MSSM. The loop diagram from the
dark matter and its partner has a negative contribution to $a_\mu$
as~\cite{Moroi:1995yh,Carena:1996qa} 
\beqa
\delta a^{(\chi,\phi)}_\mu = - \frac{\lambda^2\,m_\mu^2}{16\pi^2\,m_\phi^2} \left[ \frac{1-6x+3x^2+2x^3-6x^2\ln{x}}{6\,(1-x)^4} \right] \,,
\eeqa
with $x\equiv m_\chi^2 / m_\phi^2$. In the region with degenerate
masses, $x=1$, the part in the brackets becomes 1/12.  We show a few contours in
the $m_\chi - m_\phi$ plane in the left panel of Fig.~\ref{fig:mu-g-2}
for the fixed Yukawa coupling $\lambda=1$. Although the fermionic dark
matter case cannot explain the $(g-2)_\mu$ anomaly, the dark matter
contribution does not dramatically increase the discrepancy for a
modest $\lambda$. 

For the complex scalar dark matter case, the loop diagram from dark
matter and its partner gives a positive contribution to $a_\mu$, which
is given by 
\beqa
\delta a^{(X,\psi)}_\mu =  \frac{\lambda^2\,m_\mu^2}{16\pi^2\,m_X^2} \left[ \frac{2 + 3 x - 6x^2+  x^3+  6x\ln{x}}{6\,(1-x)^4} \right] \,,
\eeqa
with $x\equiv m_X^2 / m_\psi^2$. In the limit of $x=1$, the value in
the bracket becomes 1/12. With a large value of the coupling
$\lambda=2.5$ and a light dark matter partner mass around 150 GeV, we
show that the $(g-2)_\mu$ anomaly can be explained by the dark matter
contribution in the right panel of Fig.~\ref{fig:mu-g-2}. However, the
LUX results have significantly constrained this $(g-2)_\mu$-favored
region.

\end{appendix}

\providecommand{\href}[2]{#2}\begingroup\raggedright\endgroup


\begin{thebibliography}{10}

\bibitem{Haber:1984rc}
H.~E. Haber and G.~L. Kane, {\it {The Search for Supersymmetry: Probing Physics
  Beyond the Standard Model}},  {\em Phys.Rept.} {\bf 117} (1985) 75--263.

\bibitem{Jungman:1995df}
G.~Jungman, M.~Kamionkowski, and K.~Griest, {\it {Supersymmetric dark matter}},
   {\em Phys.Rept.} {\bf 267} (1996) 195--373,
  [\href{http://xxx.lanl.gov/abs/hep-ph/9506380}{{\tt hep-ph/9506380}}].

\bibitem{CahillRowley:2012kx}
M.~W. Cahill-Rowley, J.~L. Hewett, A.~Ismail, and T.~G. Rizzo, {\it {More
  Energy, More Searches, but the pMSSM Lives On}},  {\em Phys.Rev.} {\bf D88}
  (2013) 035002, [\href{http://xxx.lanl.gov/abs/1211.1981}{{\tt
  arXiv:1211.1981}}].

\bibitem{Cahill-Rowley:2013dpa}
M.~Cahill-Rowley, R.~Cotta, A.~Drlica-Wagner, S.~Funk, J.~Hewett, {\em
  et.~al.}, {\it {Complementarity and Searches for Dark Matter in the pMSSM}},
  \href{http://xxx.lanl.gov/abs/1305.6921}{{\tt arXiv:1305.6921}}.

\bibitem{Kolb:1990vq}
E.~W. Kolb and M.~S. Turner, {\it {The Early Universe}},  {\em Front.Phys.}
  {\bf 69} (1990) 1--547.

\bibitem{Chang:2013oia}
S.~Chang, R.~Edezhath, J.~Hutchinson, and M.~Luty, {\it {Effective WIMPs}},
  {\em Phys.Rev.} {\bf D89} (2014) 015011,
  [\href{http://xxx.lanl.gov/abs/1307.8120}{{\tt arXiv:1307.8120}}].

\bibitem{An:2013xka}
H.~An, L.-T. Wang, and H.~Zhang, {\it {Dark matter with $t$-channel mediator: a
  simple step beyond contact interaction}},
  \href{http://xxx.lanl.gov/abs/1308.0592}{{\tt arXiv:1308.0592}}.

\bibitem{Bai:2013iqa}
Y.~Bai and J.~Berger, {\it {Fermion Portal Dark Matter}},  {\em JHEP} {\bf
  1311} (2013) 171, [\href{http://xxx.lanl.gov/abs/1308.0612}{{\tt
  arXiv:1308.0612}}].

\bibitem{DiFranzo:2013vra}
A.~DiFranzo, K.~I. Nagao, A.~Rajaraman, and T.~M.~P. Tait, {\it {Simplified
  Models for Dark Matter Interacting with Quarks}},  {\em JHEP} {\bf 1311}
  (2013) 014, [\href{http://xxx.lanl.gov/abs/1308.2679}{{\tt
  arXiv:1308.2679}}].

\bibitem{Buchmueller:2013dya}
O.~Buchmueller, M.~J. Dolan, and C.~McCabe, {\it {Beyond Effective Field Theory
  for Dark Matter Searches at the LHC}},  {\em JHEP} {\bf 1401} (2014) 025,
  [\href{http://xxx.lanl.gov/abs/1308.6799}{{\tt arXiv:1308.6799}}].

\bibitem{Cheung:2013dua}
C.~Cheung and D.~Sanford, {\it {Simplified Models of Mixed Dark Matter}},
  \href{http://xxx.lanl.gov/abs/1311.5896}{{\tt arXiv:1311.5896}}.

\bibitem{Papucci:2014iwa}
M.~Papucci, A.~Vichi, and K.~M. Zurek, {\it {Monojet versus rest of the world
  I: t-channel Models}},  \href{http://xxx.lanl.gov/abs/1402.2285}{{\tt
  arXiv:1402.2285}}.

\bibitem{Simone:2014}
A.~De~Simone, G.~F. Giudice, and A.~Strumia, {\it {Benchmarks for Dark Matter
  Searches at the LHC}},  \href{http://xxx.lanl.gov/abs/1402.6287}{{\tt
  arXiv:1402.6287}}.

\bibitem{Akerib:2013tjd}
{\bf LUX} Collaboration, D.~Akerib {\em et.~al.}, {\it {First results from the
  LUX dark matter experiment at the Sanford Underground Research Facility}},
  \href{http://xxx.lanl.gov/abs/1310.8214}{{\tt arXiv:1310.8214}}.

\bibitem{Aguilar:2013qda}
{\bf AMS} Collaboration, M.~Aguilar {\em et.~al.}, {\it {First Result from the
  Alpha Magnetic Spectrometer on the International Space Station: Precision
  Measurement of the Positron Fraction in Primary Cosmic Rays of 0.5?350 GeV}},
   {\em Phys.Rev.Lett.} {\bf 110} (2013), no.~14 141102.

\bibitem{AMS-electron-positron}
{\it Positron+electron spectrum from 0.5 gev to 700 gev},  tech. rep., AMS,
  ICRC 2013.

\bibitem{Barger:2008qd}
V.~Barger, W.-Y. Keung, and G.~Shaughnessy, {\it {Spin Dependence of Dark
  Matter Scattering}},  {\em Phys.Rev.} {\bf D78} (2008) 056007,
  [\href{http://xxx.lanl.gov/abs/0806.1962}{{\tt arXiv:0806.1962}}].

\bibitem{Batell:2013zwa}
B.~Batell, T.~Lin, and L.-T. Wang, {\it {Flavored Dark Matter and R-Parity
  Violation}},  \href{http://xxx.lanl.gov/abs/1309.4462}{{\tt
  arXiv:1309.4462}}.

\bibitem{Ellis:1999mm}
J.~R. Ellis, T.~Falk, K.~A. Olive, and M.~Srednicki, {\it {Calculations of
  neutralino-stau coannihilation channels and the cosmologically relevant
  region of MSSM parameter space}},  {\em Astropart.Phys.} {\bf 13} (2000)
  181--213, [\href{http://xxx.lanl.gov/abs/hep-ph/9905481}{{\tt
  hep-ph/9905481}}].

\bibitem{Arnowitt:2008bz}
R.~L. Arnowitt, B.~Dutta, A.~Gurrola, T.~Kamon, A.~Krislock, {\em et.~al.},
  {\it {Determining the Dark Matter Relic Density in the mSUGRA (~X0(1))-~tau
  Co-Annhiliation Region at the LHC}},  {\em Phys.Rev.Lett.} {\bf 100} (2008)
  231802, [\href{http://xxx.lanl.gov/abs/0802.2968}{{\tt arXiv:0802.2968}}].

\bibitem{Kopp:2009et}
J.~Kopp, V.~Niro, T.~Schwetz, and J.~Zupan, {\it {DAMA/LIBRA and leptonically
  interacting Dark Matter}},  {\em Phys.Rev.} {\bf D80} (2009) 083502,
  [\href{http://xxx.lanl.gov/abs/0907.3159}{{\tt arXiv:0907.3159}}].

\bibitem{Agrawal:2011ze}
P.~Agrawal, S.~Blanchet, Z.~Chacko, and C.~Kilic, {\it {Flavored Dark Matter,
  and Its Implications for Direct Detection and Colliders}},  {\em Phys.Rev.}
  {\bf D86} (2012) 055002, [\href{http://xxx.lanl.gov/abs/1109.3516}{{\tt
  arXiv:1109.3516}}].

\bibitem{Fitzpatrick:2010br}
A.~L. Fitzpatrick and K.~M. Zurek, {\it {Dark Moments and the DAMA-CoGeNT
  Puzzle}},  {\em Phys.Rev.} {\bf D82} (2010) 075004,
  [\href{http://xxx.lanl.gov/abs/1007.5325}{{\tt arXiv:1007.5325}}].

\bibitem{Ho:2012bg}
C.~M. Ho and R.~J. Scherrer, {\it {Anapole Dark Matter}},  {\em Phys.Lett.}
  {\bf B722} (2013) 341--346, [\href{http://xxx.lanl.gov/abs/1211.0503}{{\tt
  arXiv:1211.0503}}].

\bibitem{DelNobile:2014eta}
E.~Del~Nobile, G.~B. Gelmini, P.~Gondolo, and J.-H. Huh, {\it {Direct detection
  of Light Anapole and Magnetic Dipole DM}},
  \href{http://xxx.lanl.gov/abs/1401.4508}{{\tt arXiv:1401.4508}}.

\bibitem{Raghavan:1989zz}
P.~Raghavan, {\it {Table of nuclear moments}},  {\em Atom.Data Nucl.Data Tabl.}
  {\bf 42} (1989) 189--291.

\bibitem{Banks:2010eh}
T.~Banks, J.-F. Fortin, and S.~Thomas, {\it {Direct Detection of Dark Matter
  Electromagnetic Dipole Moments}},
  \href{http://xxx.lanl.gov/abs/1007.5515}{{\tt arXiv:1007.5515}}.

\bibitem{Consortium:2010bc}
{\bf CTA Consortium} Collaboration, M.~Actis {\em et.~al.}, {\it {Design
  concepts for the Cherenkov Telescope Array CTA: An advanced facility for
  ground-based high-energy gamma-ray astronomy}},  {\em Exper.Astron.} {\bf 32}
  (2011) 193--316, [\href{http://xxx.lanl.gov/abs/1008.3703}{{\tt
  arXiv:1008.3703}}].

\bibitem{Garny:2012eb}
M.~Garny, A.~Ibarra, M.~Pato, and S.~Vogl, {\it {Closing in on mass-degenerate
  dark matter scenarios with antiprotons and direct detection}},  {\em JCAP}
  {\bf 1211} (2012) 017, [\href{http://xxx.lanl.gov/abs/1207.1431}{{\tt
  arXiv:1207.1431}}].

\bibitem{Garny:2013ama}
M.~Garny, A.~Ibarra, M.~Pato, and S.~Vogl, {\it {Internal bremsstrahlung
  signatures in light of direct dark matter searches}},  {\em JCAP} {\bf 1312}
  (2013) 046, [\href{http://xxx.lanl.gov/abs/1306.6342}{{\tt
  arXiv:1306.6342}}].

\bibitem{Adriani:2008zr}
{\bf PAMELA} Collaboration, O.~Adriani {\em et.~al.}, {\it {An anomalous
  positron abundance in cosmic rays with energies 1.5-100 GeV}},  {\em Nature}
  {\bf 458} (2009) 607--609, [\href{http://xxx.lanl.gov/abs/0810.4995}{{\tt
  arXiv:0810.4995}}].

\bibitem{FermiLAT:2011ab}
{\bf Fermi LAT} Collaboration, M.~Ackermann {\em et.~al.}, {\it {Measurement of
  separate cosmic-ray electron and positron spectra with the Fermi Large Area
  Telescope}},  {\em Phys.Rev.Lett.} {\bf 108} (2012) 011103,
  [\href{http://xxx.lanl.gov/abs/1109.0521}{{\tt arXiv:1109.0521}}].

\bibitem{Cholis:2013psa}
I.~Cholis and D.~Hooper, {\it {Dark Matter and Pulsar Origins of the Rising
  Cosmic Ray Positron Fraction in Light of New Data From AMS}},  {\em
  Phys.Rev.} {\bf D88} (2013) 023013,
  [\href{http://xxx.lanl.gov/abs/1304.1840}{{\tt arXiv:1304.1840}}].

\bibitem{Cirelli:2008id}
M.~Cirelli, R.~Franceschini, and A.~Strumia, {\it {Minimal Dark Matter
  predictions for galactic positrons, anti-protons, photons}},  {\em
  Nucl.Phys.} {\bf B800} (2008) 204--220,
  [\href{http://xxx.lanl.gov/abs/0802.3378}{{\tt arXiv:0802.3378}}].

\bibitem{Bai:2009ka}
Y.~Bai, M.~Carena, and J.~Lykken, {\it {The PAMELA excess from neutralino
  annihilation in the NMSSM}},  {\em Phys.Rev.} {\bf D80} (2009) 055004,
  [\href{http://xxx.lanl.gov/abs/0905.2964}{{\tt arXiv:0905.2964}}].

\bibitem{Sjostrand:2007gs}
T.~Sjostrand, S.~Mrenna, and P.~Z. Skands, {\it {A Brief Introduction to PYTHIA
  8.1}},  {\em Comput.Phys.Commun.} {\bf 178} (2008) 852--867,
  [\href{http://xxx.lanl.gov/abs/0710.3820}{{\tt arXiv:0710.3820}}].

\bibitem{Delahaye:2007fr}
T.~Delahaye, R.~Lineros, F.~Donato, N.~Fornengo, and P.~Salati, {\it {Positrons
  from dark matter annihilation in the galactic halo: Theoretical
  uncertainties}},  {\em Phys.Rev.} {\bf D77} (2008) 063527,
  [\href{http://xxx.lanl.gov/abs/0712.2312}{{\tt arXiv:0712.2312}}].

\bibitem{Bergstrom:2013jra}
L.~Bergstrom, T.~Bringmann, I.~Cholis, D.~Hooper, and C.~Weniger, {\it {New
  limits on dark matter annihilation from AMS cosmic ray positron data}},  {\em
  Phys.Rev.Lett.} {\bf 111} (2013) 171101,
  [\href{http://xxx.lanl.gov/abs/1306.3983}{{\tt arXiv:1306.3983}}].

\bibitem{Ibarra:2013zia}
A.~Ibarra, A.~S. Lamperstorfer, and J.~Silk, {\it {Dark matter annihilations
  and decays after the AMS-02 positron measurements}},
  \href{http://xxx.lanl.gov/abs/1309.2570}{{\tt arXiv:1309.2570}}.

\bibitem{Beenakker:1996ed}
W.~Beenakker, R.~Hopker, and M.~Spira, {\it {PROSPINO: A Program for the
  production of supersymmetric particles in next-to-leading order QCD}},
  \href{http://xxx.lanl.gov/abs/hep-ph/9611232}{{\tt hep-ph/9611232}}.

\bibitem{Fuks:2013lya}
B.~Fuks, M.~Klasen, D.~R. Lamprea, and M.~Rothering, {\it {Revisiting slepton
  pair production at the Large Hadron Collider}},
  \href{http://xxx.lanl.gov/abs/1310.2621}{{\tt arXiv:1310.2621}}.

\bibitem{ATLAS-CONF-2013-049}
{\bf ATLAS} Collaboration, {\it {Search for direct-slepton and direct-chargino
  production in final states with two opposite-sign leptons, missing transverse
  momentum and no jets in 20/fb of pp collisions at sqrt(s) = 8 TeV with the
  ATLAS detector}},  Tech. Rep. ATLAS-CONF-2013-049, CERN, Geneva, May, 2013.

\bibitem{CMS-PAS-SUS-13-006}
{\bf CMS} Collaboration, {\it {Search for electroweak production of charginos,
  neutralinos, and sleptons using leptonic final states in pp collisions at 8
  TeV}},  Tech. Rep. CMS-PAS-SUS-13-006, CERN, Geneva, 2013.

\bibitem{Matchev:2009ad}
K.~T. Matchev and M.~Park, {\it {A General method for determining the masses of
  semi-invisibly decaying particles at hadron colliders}},  {\em
  Phys.Rev.Lett.} {\bf 107} (2011) 061801,
  [\href{http://xxx.lanl.gov/abs/0910.1584}{{\tt arXiv:0910.1584}}].

\bibitem{Tovey:2008ui}
D.~R. Tovey, {\it {On measuring the masses of pair-produced semi-invisibly
  decaying particles at hadron colliders}},  {\em JHEP} {\bf 0804} (2008) 034,
  [\href{http://xxx.lanl.gov/abs/0802.2879}{{\tt arXiv:0802.2879}}].

\bibitem{Buckley:2013kua}
M.~R. Buckley, J.~D. Lykken, C.~Rogan, and M.~Spiropulu, {\it {Super-Razor and
  Searches for Sleptons and Charginos at the LHC}},
  \href{http://xxx.lanl.gov/abs/1310.4827}{{\tt arXiv:1310.4827}}.

\bibitem{Lester:1999tx}
C.~Lester and D.~Summers, {\it {Measuring masses of semiinvisibly decaying
  particles pair produced at hadron colliders}},  {\em Phys.Lett.} {\bf B463}
  (1999) 99--103, [\href{http://xxx.lanl.gov/abs/hep-ph/9906349}{{\tt
  hep-ph/9906349}}].

\bibitem{Barr:2003rg}
A.~Barr, C.~Lester, and P.~Stephens, {\it {m(T2): The Truth behind the
  glamour}},  {\em J.Phys.} {\bf G29} (2003) 2343--2363,
  [\href{http://xxx.lanl.gov/abs/hep-ph/0304226}{{\tt hep-ph/0304226}}].

\bibitem{Cheng:2008hk}
H.-C. Cheng and Z.~Han, {\it {Minimal Kinematic Constraints and m(T2)}},  {\em
  JHEP} {\bf 0812} (2008) 063, [\href{http://xxx.lanl.gov/abs/0810.5178}{{\tt
  arXiv:0810.5178}}].

\bibitem{Konar:2009qr}
P.~Konar, K.~Kong, K.~T. Matchev, and M.~Park, {\it {Dark Matter Particle
  Spectroscopy at the LHC: Generalizing M(T2) to Asymmetric Event Topologies}},
   {\em JHEP} {\bf 1004} (2010) 086,
  [\href{http://xxx.lanl.gov/abs/0911.4126}{{\tt arXiv:0911.4126}}].

\bibitem{Kats:2011qh}
Y.~Kats, P.~Meade, M.~Reece, and D.~Shih, {\it {The Status of GMSB After 1/fb
  at the LHC}},  {\em JHEP} {\bf 1202} (2012) 115,
  [\href{http://xxx.lanl.gov/abs/1110.6444}{{\tt arXiv:1110.6444}}].

\bibitem{Bai:2012gs}
Y.~Bai, H.-C. Cheng, J.~Gallicchio, and J.~Gu, {\it {Stop the Top Background of
  the Stop Search}},  {\em JHEP} {\bf 1207} (2012) 110,
  [\href{http://xxx.lanl.gov/abs/1203.4813}{{\tt arXiv:1203.4813}}].

\bibitem{Kilic:2012kw}
C.~Kilic and B.~Tweedie, {\it {Cornering Light Stops with Dileptonic mT2}},
  {\em JHEP} {\bf 1304} (2013) 110,
  [\href{http://xxx.lanl.gov/abs/1211.6106}{{\tt arXiv:1211.6106}}].

\bibitem{Bai:2013ema}
Y.~Bai, H.-C. Cheng, J.~Gallicchio, and J.~Gu, {\it {A Toolkit of the Stop
  Search via the Chargino Decay}},  {\em JHEP} {\bf 1308} (2013) 085,
  [\href{http://xxx.lanl.gov/abs/1304.3148}{{\tt arXiv:1304.3148}}].

\bibitem{Arnison:1983rp}
{\bf UA1} Collaboration, G.~Arnison {\em et.~al.}, {\it {Experimental
  Observation of Isolated Large Transverse Energy Electrons with Associated
  Missing Energy at s**(1/2) = 540-GeV}},  {\em Phys.Lett.} {\bf B122} (1983)
  103--116.

\bibitem{vanNeerven:1982mz}
W.~van Neerven, J.~Vermaseren, and K.~Gaemers, {\it {LEPTON - JET EVENTS AS A
  SIGNATURE FOR W PRODUCTION IN p anti-p COLLISIONS}}, .

\bibitem{Barger:1983wf}
V.~D. Barger, A.~D. Martin, and R.~Phillips, {\it {Perpendicular $\nu_e$ Mass
  From $W$ Decay}},  {\em Z.Phys.} {\bf C21} (1983) 99.

\bibitem{Smith:1983aa}
J.~Smith, W.~van Neerven, and J.~Vermaseren, {\it {The Transverse Mass and
  Width of the $W$ Boson}},  {\em Phys.Rev.Lett.} {\bf 50} (1983) 1738.

\bibitem{Alwall:2011uj}
J.~Alwall, M.~Herquet, F.~Maltoni, O.~Mattelaer, and T.~Stelzer, {\it {MadGraph
  5 : Going Beyond}},  {\em JHEP} {\bf 1106} (2011) 128,
  [\href{http://xxx.lanl.gov/abs/1106.0522}{{\tt arXiv:1106.0522}}].

\bibitem{Christensen:2008py}
N.~D. Christensen and C.~Duhr, {\it {FeynRules - Feynman rules made easy}},
  {\em Comput.Phys.Commun.} {\bf 180} (2009) 1614--1641,
  [\href{http://xxx.lanl.gov/abs/0806.4194}{{\tt arXiv:0806.4194}}].

\bibitem{PGS}
J.~S. Conway, {\it {Pretty Good Simulation of high-energy collisions}},
  \href{http://xxx.lanl.gov/abs/090401 release}{{\tt 090401 release}}.

\bibitem{Abazov:2009vs}
{\bf D0} Collaboration, V.~Abazov {\em et.~al.}, {\it {Direct measurement of
  the W boson width}},  {\em Phys.Rev.Lett.} {\bf 103} (2009) 231802,
  [\href{http://xxx.lanl.gov/abs/0909.4814}{{\tt arXiv:0909.4814}}].

\bibitem{Hagiwara:2011af}
K.~Hagiwara, R.~Liao, A.~D. Martin, D.~Nomura, and T.~Teubner, {\it
  {$(g-2)_\mu$ and $\alpha(M_Z^2)$ re-evaluated using new precise data}},  {\em
  J.Phys.} {\bf G38} (2011) 085003,
  [\href{http://xxx.lanl.gov/abs/1105.3149}{{\tt arXiv:1105.3149}}].

\bibitem{Bennett:2006fi}
{\bf Muon G-2} Collaboration, G.~Bennett {\em et.~al.}, {\it {Final Report of
  the Muon E821 Anomalous Magnetic Moment Measurement at BNL}},  {\em
  Phys.Rev.} {\bf D73} (2006) 072003,
  [\href{http://xxx.lanl.gov/abs/hep-ex/0602035}{{\tt hep-ex/0602035}}].

\bibitem{Roberts:2010cj}
B.~L. Roberts, {\it {Status of the Fermilab Muon $(g-2)$ Experiment}},  {\em
  Chin.Phys.} {\bf C34} (2010) 741--744,
  [\href{http://xxx.lanl.gov/abs/1001.2898}{{\tt arXiv:1001.2898}}].

\bibitem{Blum:2013xva}
T.~Blum, A.~Denig, I.~Logashenko, E.~de~Rafael, B.~Lee~Roberts, {\em et.~al.},
  {\it {The Muon (g-2) Theory Value: Present and Future}},
  \href{http://xxx.lanl.gov/abs/1311.2198}{{\tt arXiv:1311.2198}}.

\bibitem{Moroi:1995yh}
T.~Moroi, {\it {The Muon anomalous magnetic dipole moment in the minimal
  supersymmetric standard model}},  {\em Phys.Rev.} {\bf D53} (1996)
  6565--6575, [\href{http://xxx.lanl.gov/abs/hep-ph/9512396}{{\tt
  hep-ph/9512396}}].

\bibitem{Carena:1996qa}
M.~S. Carena, G.~Giudice, and C.~Wagner, {\it {Constraints on supersymmetric
  models from the muon anomalous magnetic moment}},  {\em Phys.Lett.} {\bf
  B390} (1997) 234--242, [\href{http://xxx.lanl.gov/abs/hep-ph/9610233}{{\tt
  hep-ph/9610233}}].

\end{thebibliography}
 \end{document}